\documentclass[prb,twocolumn,aps,showpacs]{revtex4}
\usepackage{graphicx}
\usepackage{bm}
\usepackage{amssymb}
\usepackage{amsmath}

\input{tcilatex.tex}

\input{tcilatex}

\begin{document}

\title{Low Energy Physical Properties of High Tc Superconducting Cu-oxides -
A Comparison Between Plain Vanilla RVB Theory and Experiments}
\author{Kai-Yu Yang$^{1}$, C.T. Shih$^{2}$, C. P. Chou$^{3}$, S. M. Huang$%
^{3} $, T. K. Lee$^{3}$, T. Xiang$^{4}$, F. C. Zhang$^{1,5}$}
\date{\today }
\affiliation{$^{1}$Centre of Theoretical and Computational Physics, and Department of
Physics, The University of Hong Kong, Hong Kong\\
$^{2}$Department of Physics, Tunghai University, Taichung, Taiwan\\
$^{3}$Institute of Physics, Academia Sinica, Taipei, Taiwan\\
$^{4}$Institute of Theoretical Physics and Interdisciplinary Center of
Theoretical Studies, Chinese Academy of Sciences, Beijing, China\\
$^5$ Department of Physics, University of Cincinnati, Cincinnati, Ohio 45221}

\begin{abstract}
In a recent review by Anderson and coworkers\cite{Vanilla}, it was pointed
out that an early resonating valence bond (RVB) theory is able to explain a
number of unusual properties of high temperature superconducting (SC)
Cu-oxides. Here we extend previous calculations \cite{anderson87,FC
Zhang,Randeria} to study more systematically the low energy physical
properties of the plain vanilla d-wave RVB state, and to compare the results
with the available experiments. We use a renormalized mean field theory
combined with variational Monte Carlo and power Lanczos methods to study the
RVB state of an extended $t-J$ model in a square lattice with parameters
suitable for the hole doped Cu-oxides. The physical observable quantities we
study include the specific heat, the linear residual thermal conductivity,
the in-plane magnetic penetration depth, the quasiparticle energy at the
antinode $(\pi ,0)$, the superconducting energy gap, the quasiparticle
spectra and the Drude weights. The traits of nodes (including $k_{F}$, the
Fermi velocity $v_{F}$ and the velocity along Fermi surface $v_{2}$), as
well as the SC order parameter are also studied. Comparisons of the theory
and the experiments in cuprates show an overall qualitative agreement,
especially on their doping dependences.
\end{abstract}

\pacs{74.20.Fg, 74.72.-h, 71.10.-w}
\maketitle

\section{INTRODUCTION}

Since the discovery of high transition temperature superconductivity
(HTSC) in cuprates in 1986, there have been enormous efforts in both
experiments and theories to understand the mechanism of the
superconductivity and their unusual physical properties. One of the
earliest theory is the resonating valence bond (RVB) theory proposed
by Anderson\cite{anderson87}. In that theory, the operative element
in the electronic structure of this class of compounds is the square
planar $CuO_{2}$ lattice. The parent compound such as
$La_{2}CuO_{4}$, where the $Cu$ is stoichiometrically bivalent
$Cu^{2+}$ with one hole per $Cu$ site, is an antiferromagnetically
coupled Mott insulator. Chemical doping such as the partial
replacement of $La$ by $Sr$ introduces additional holes on the
$CuO_{2}$ plane. The minimum microscopic model to describe the low
energy physics has been proposed to be the $t-J$ model or its
variant extended $t-J$ model, which includes an antiferromagnetic
spin coupling and a kinetic energy term for the hole motion
\cite{anderson87,Zhang-Rice}. Anderson proposed a doped spin liquid
of spin singlets, or the bond spin singlet resonating between many
configurations.\cite{anderson87} This concept explains many unusual
properties of the cuprates, as emphasized in a recent review by
Anderson and coworkers\cite{Vanilla}. More quantitatively, in the
simplest RVB theory, namely its plain vanilla version, the RVB state
in the cuprate is described by a Gutzwiller projected d-wave BCS
wavefunction, whose parameters are determined variationally either
by using a renormalized mean field theory (RMFT)\cite{FC Zhang} or
by variational Monte Carlo method (VMC) numerically
\cite{Gros89,TKLee,CTShih}, or by other field theory
methods\cite{Kotliar}. Recently the Gutzwiller RVB wavefunction
approach was applied to the strong coupling Hubbard model by
Paramekanti, Randeria and Trivedi\cite{Randeria}, who used careful
numerical methods to calculate several quantities of direct
experimental relevance. Both results for the Hubbard and $t-J$
models turn out to correspond well to some experimental phenomena
observed in cuprates. The plain vanilla RVB theory has recently been
extended to study the scanning tunneling microscopy, the angle
resolved photoemission spectroscopy (ARPES) \cite{anderson-ong} and
the Gossamer superconductivity\cite{Zhang03} in the Hubbard-like
models at the half electron filling. In view of the preliminary
success of the plain vanilla RVB theory, it is desirable to extend
previous calculations and to analyze more experimental data so that
a
more systematic and comprehensive comparison between the theories \cite%
{BEdegger,Rice06} and the experiments \cite{TYoshida} can be made on more
observable quantities.

In the present paper, we extend the previous works of Zhang et al.
\cite{FC Zhang} and of Paramekanti et al. \cite{Randeria} to carry
out more systematic calculations on the low energy physical
properties of the plain vanilla d-wave RVB state. We use a
renormalized mean field theory combined with variational Monte Carlo
and power Lanczos methods (PL)\cite{Y.C. Chen,C.T. Shih} to study
the RVB state of an extended $t-J$ model in a square lattice for
parameters suitable for the hole doped lanthanum and yttrium
Cu-oxides. Our main focus is on the microscopic calculations of the
key parameters for the nodal quasiparticles in the d-wave RVB state,
namely the Fermi velocity $v_{F}$ and the velocity along the Fermi
surface $v_{2}$. From these quantities, we calculate a number of low
energy physical properties including the specific heat, the linear
residual thermal conductivity, and the in-plane magnetic penetration
depth. We make extensive comparison between these calculations with
a very broad spectrum of types of reported experiments, and find
qualitative agreement, especially on the doping dependences of these
properties. The discrepancy between the theory and the experiments
is mostly on the absolute values of these quantities, which may be
attributed to a factor of $2\sim 4$ times larger in the value of
$v_{2}$ in the theory. We also calculate the quasiparticle energy
gap at the antinode $(\pi ,0)$, the SC energy gap, the quasiparticle
spectra and the Drude weight, and find good agreements with the
experiments.

The paper is organized as follows. In section II, we describe the
microscopic model and the methods we use in our calculations. In section
III, we calculate the basic parameters of the nodal quasiparticles. In
section IV, we discuss the nodal physics and make comparison of the theory
and experiments on a number of low energy physical properties. In section V,
we calculate other physical quantities and compare with the experiments. A
brief summary is given in section VI.

\section{MODEL AND METHODS}

\subsection{model}

We consider an extended $t-J$ model including a nearest neighbor (n.n.), a
second n.n. and a third n.n. hopping terms in a square lattice,

\begin{eqnarray}
H &=&H_{t}+H_{J}  \label{Hamiltonian} \\
H_{t} &=&-t\underset{\left\langle i,j\right\rangle ,\sigma }{\sum }%
c_{i\sigma }^{\dagger }c_{j\sigma }-t^{\prime }\underset{\left\langle
i,j\right\rangle ^{^{\prime }},\sigma }{\sum }c_{i\sigma }^{\dagger
}c_{j\sigma }-t^{\prime \prime }\underset{\left\langle i,j\right\rangle
^{\prime \prime },\sigma }{\sum }c_{i\sigma }^{\dagger }c_{j\sigma }  \notag
\\
H_{J} &=&J\underset{\left\langle i,j\right\rangle }{\sum }\boldsymbol{S}%
_{i}\cdot \boldsymbol{S}_{j}  \notag
\end{eqnarray}%
In the above Hamiltonian, a constraint of no double occupation of electrons
on each site is implied: $\sum_{\sigma }c_{i,\sigma }^{\dagger }c_{i,\sigma
}\leq 1$. The summations $\left\langle i,j\right\rangle ,\left\langle
i,j\right\rangle ^{\prime }$ and $\left\langle i,j\right\rangle ^{\prime
\prime }$ run over the n.n., second n.n. and third n.n. pairs, respectively.
$t$, $t^{\prime }$ and $t^{\prime \prime }$ are their corresponding hopping
integrals respectively. We choose $t$ and $t^{\prime \prime }$ to be
positive and $t^{\prime }$ to be negative, appropriate for the hole-doped
cuprates; $J$ is the superexchange coupling between the n.n. spins. Table I
lists the parameters $\{t,t^{\prime }/t,t^{\prime \prime }/t,J/t\}$ used in
our calculations for mono-layered $La_{2-x}Sr_{x}CuO_{4}$ (LSCO) and
bi-layered cuprate $YBa_{2}Cu_{3}O_{7-x}$ (YBCO)\ or $%
Bi_{2}Sr_{2}CaCu_{2}O_{8+x}$ (Bi-2212). These parameters appear consistent
with the band-structure calculations \cite{E. Pavarini} and also with the
experimental analyses such as the topology of large Fermi surface reported
in ARPES \cite{Z.X. Shen, A. Ino2}, the inelastic light scattering \cite{K.
B. Lyons}, neutron scattering \cite{J. M. Tranquada, R. Coldea, S. M. Hayden}
and two-magnon Raman scattering \cite{G. Blumberg, S. Sugai, G. Blumberg1}
experiments.

\bigskip

\bigskip TABLE I:\ Parameters $\{t,t^{\prime }/t,t^{\prime \prime }/t,J/t\}$
used for $\ La_{2-x}Sr_{x}CuO_{4}$ and $YBa_{2}Cu_{3}O_{7-x}$ $%
(Bi_{2}Sr_{2}CaCu_{2}O_{8+x})$\ in the renormalized mean field theory.
\bigskip

\begin{tabular}{ccccc}
\hline\hline
& $\ \ t(eV)$ & $\ \ t^{\prime }/t$ & $\ t^{\prime \prime }/t$ & $\ \ J/t$ \
\  \\ \hline
$\ \ \ \ \ La_{2-x}Sr_{x}CuO_{4}$ \ \ \ \ \  & \ $0.3$ & \ \ $-0.1$ & \ $%
0.05 $ & \ $0.3$ \  \\ \hline
\begin{tabular}{l}
$\ \ YBa_{2}Cu_{3}O_{7-x}$ \\
$\ Bi_{2}Sr_{2}CaCu_{2}O_{8+x}$%
\end{tabular}
& \ $0.3$ & \ $-0.3$ & \ $0.2$ & \ $0.3$ \  \\ \hline\hline
\end{tabular}

\medskip \bigskip \bigskip

We use a variational projected $d$-wave BCS state or the d-RVB state to
study the ground state and elementary excitations of the model. \cite%
{anderson87} The trial ground state is of the form,
\begin{equation}
\left\vert \Psi _{L}\right\rangle =P_{G}\left\vert \Psi _{BCS}\right\rangle ,
\label{wavefunction}
\end{equation}%
where the Gutzwiller projection operator $P_{G}=\prod_{i}(1-n_{i\uparrow
}n_{i\downarrow })$ is to ensure the constraint of no double occupation of
electrons on any lattice site. The BCS state is of the standard form, given
by
\begin{equation*}
\left\vert \Psi _{BCS}\right\rangle =\underset{\boldsymbol{k}}{\prod }(u_{%
\boldsymbol{k}}+v_{\boldsymbol{k}}c_{\boldsymbol{k}\uparrow }^{\dagger }c_{-%
\boldsymbol{k}\downarrow }^{\dagger })\left\vert 0\right\rangle
\end{equation*}%
where $\left\vert 0\right\rangle $ is the vacuum, and $u_{\boldsymbol{k}}$
and $v_{\boldsymbol{k}}$ are the variational parameters satisfying the
normalization condition: $\left\vert u_{\boldsymbol{k}}\right\vert
^{2}+\left\vert v_{\boldsymbol{k}}\right\vert ^{2}=1$.

In this paper, we use two complementary methods to carry out the Gutzwiller
projected variational calculation. One is the renormalized mean-field theory
(RMFT), which takes into account of the Gutzwiller projection by a set of
renormalization factors. \cite{FC Zhang} The other is the variational Monte
Carlo (VMC) method which computes the quantities numerically, followed by a
further improvement of the variational wavefunction by using the power
Lanczos (PL) method \cite{Y.C. Chen,C.T. Shih} to eliminate or to reduce the
bias in the variational approach. It is well known that the variational
calculation often overestimates the effect of superconductivity in the true
ground state, and the variational calculation usually leads to a larger $%
\Delta $.\cite{CTShih}

\subsection{Renormalized mean field theory}

The RMFT is a Hartree-Fock like mean field theory to approximately treat the
projection operator in the Hamiltonian (\ref{Hamiltonian}). In the RMFT, we
apply the Gutzwiller approximation to replace the effect of the projection
operator by a set of renormalization factors, which are determined by
statistical counting \cite{FC Zhang, Gutzwiller}. The variation of a
projected state for Hamiltonian $H$ is then approximately mapped onto that
of the corresponding unprojected state for a renormalized Hamiltonian \cite%
{FC Zhang, Vanilla}. This method was initially developed by Gutzwiller to
study possible ferromagnetism in strongly interacting systems \cite%
{Gutzwiller}. It was later applied by Brinkman and Rice to study the metal
insulator transition~\cite{Brinkman-rice}, and by Vollhardt to study the
Fermi liquid theory of Helium-3 \cite{Vollhardt}.

Let $\langle Q\rangle $ be an expectation value of $Q$ in the RVB state $%
\left\vert \Psi _{L}\right\rangle $, and $\langle Q\rangle _{0}$ be an
expectation value of $Q$ in the BCS state $\left\vert \Psi
_{BCS}\right\rangle $, then the expectation values of the hopping term and
the spin-spin correlation in the RVB states can be written in terms of those
in the BCS state,
\begin{eqnarray}
\langle c_{i\sigma }^{\dagger }c_{j\sigma }\rangle  &=&g_{t}\langle
c_{i\sigma }^{\dagger }c_{j\sigma }\rangle _{0},  \notag \\
\langle \boldsymbol{S}_{i}\cdot \boldsymbol{S}_{j}\rangle  &=&g_{s}\langle
\boldsymbol{S}_{i}\cdot \boldsymbol{S}_{j}\rangle _{0}  \label{refactor}
\end{eqnarray}%
where $g_{t}$ and $g_{s}$ are the two renormalization factors for the
kinetic and the spin--spin superexchange terms respectively, they are given
by \cite{FC Zhang}
\begin{equation*}
g_{t}=\frac{2\delta }{1+\delta },\,\,g_{s}=\frac{4}{(1+\delta )^{2}}
\end{equation*}%
with $\delta $ the hole concentration. The evaluation of $H$ in the RVB
state is then mapped onto the evaluation of the renormalized Hamiltonian $%
H^{\prime }$ in the corresponding BCS state, with $H^{\prime }$ given by%
\begin{equation}
H^{\prime }=g_{t}H_{t}+g_{s}H_{J}  \label{rehamiltonian}
\end{equation}%
The variational energy of the system is then given by
\begin{equation*}
W=\langle H\rangle =\langle H^{\prime }\rangle _{0}
\end{equation*}%
In this paper, we shall only consider even parity SC state, namely $|v_{-%
\boldsymbol{k}}|^{2}=|v_{\boldsymbol{k}}|^{2}$, and $v_{\boldsymbol{k}%
}^{\ast }u_{\boldsymbol{k}}=u_{-\boldsymbol{k}}^{\ast }v_{-\boldsymbol{k}}$.
We obtain
\begin{eqnarray*}
W &=&2g_{t}\sum_{\boldsymbol{k}}|v_{\boldsymbol{k}}|^{2}\varepsilon (%
\boldsymbol{k}) \\
&&+\frac{g_{s}}{N}\sum_{\boldsymbol{k},\boldsymbol{k}^{\prime }}V_{%
\boldsymbol{k}-\boldsymbol{k}^{\prime }}(|v_{\boldsymbol{k}}|^{2}|v_{%
\boldsymbol{k}^{\prime }}|^{2}+u_{\boldsymbol{k}}v_{\boldsymbol{k}}v_{%
\boldsymbol{k}^{\prime }}^{\ast }u_{\boldsymbol{k}^{\prime }}^{\ast })
\end{eqnarray*}%
where
\begin{eqnarray*}
\varepsilon (\boldsymbol{k}) &=&-2t(\cos k_{x}+\cos k_{y})-4t^{\prime }(\cos
k_{x}\cos k_{y}) \\
&&-2t^{\prime \prime }(\cos 2k_{x}+\cos 2k_{y}), \\
V_{\boldsymbol{k}-\boldsymbol{k}^{\prime }} &=&-\frac{3}{2}J[\cos
(k_{x}-k_{x}^{\prime })+\cos (k_{y}-k_{y}^{\prime })]
\end{eqnarray*}%
with $N$ the total number of lattice sites. The total number of electron
operator $N_{e}=\sum_{\boldsymbol{k}\sigma }c_{\boldsymbol{k}\sigma }^{\dag
}c_{\boldsymbol{k}\sigma }$ has an expectation value of $\left\langle
N_{e}\right\rangle =2\sum_{\boldsymbol{k}}|v_{\boldsymbol{k}}|^{2}$, so that
the hole concentration
\begin{equation}
\delta =1-2\sum_{\boldsymbol{k}}|v_{\boldsymbol{k}}|^{2}/N
\label{hole_concentration}
\end{equation}%
Let $\mu $ be the chemical potential, the quantity we wish to minimize is $%
W^{\prime }=\left\langle H-\mu N_{e}\right\rangle $, or
\begin{equation*}
W^{\prime }=W-2\mu \sum_{\boldsymbol{k}}|v_{\boldsymbol{k}}|^{2}
\end{equation*}%
The variation is carried out respect to $v_{\boldsymbol{k}},u_{\boldsymbol{k}%
}$ and $\delta $ for fixed $\mu $. Carrying out this variational procedure,
we find that
\begin{eqnarray}
|v_{\boldsymbol{k}}|^{2} &=&\frac{1}{2}(1-\xi (\boldsymbol{k})/E(\boldsymbol{%
k})),  \notag \\
|u_{\boldsymbol{k}}|^{2} &=&\frac{1}{2}(1+\xi (\boldsymbol{k})/E(\boldsymbol{%
k})),  \notag \\
u_{\boldsymbol{k}}v_{\boldsymbol{k}} &=&\Delta (\boldsymbol{k})/2E(%
\boldsymbol{k})  \label{self-consistent}
\end{eqnarray}%
with
\begin{equation*}
E(\boldsymbol{k})=\sqrt{\xi ^{2}(\boldsymbol{k})+\left\vert \Delta (%
\boldsymbol{k})\right\vert ^{2}}
\end{equation*}%
The parameters $\xi (\boldsymbol{k})$ and $\Delta (\boldsymbol{k})$ are
related to the particle-hole and particle-particle pairing amplitudes which
are introduced below in Eq.[\ref{particle-particle},\ref{particle-hole}], $E(%
\boldsymbol{k})$ turns out to be the energy of a quasiparticle in the SC
state. \cite{FC Zhang} We define
\begin{eqnarray}
\Delta _{\tau } &=&\langle c_{i\uparrow }^{\dag }c_{i+\tau \downarrow
}^{\dag }-c_{i\downarrow }^{\dag }c_{i+\tau \uparrow }^{\dag }\rangle ,
\label{particle-particle} \\
\chi _{\tau } &=&\sum_{\sigma }\langle c_{i\sigma }^{\dag }c_{i+\tau \sigma
}\rangle   \label{particle-hole}
\end{eqnarray}%
with $\tau =\hat{x},\hat{y}$, the n.n. unit vector. For the $d_{x^{2}-y^{2}}$
pairing symmetry, $\Delta _{x}=-\Delta _{y}=\Delta _{0}$, $\chi _{x}=\chi
_{y}=\chi _{0}$, and $\xi (\boldsymbol{k})$ and $\Delta (\boldsymbol{k})$
have the forms
\begin{eqnarray}
\xi (\boldsymbol{k}) &=&g_{t}\varepsilon (\boldsymbol{k})-\widetilde{\mu }%
-\chi (\cos k_{x}+\cos k_{y}),  \label{dispersion1} \\
\Delta (\boldsymbol{k}) &=&\Delta (\cos k_{x}-\cos k_{y})
\label{dispersion2}
\end{eqnarray}%
where $\Delta =(3g_{s}J/4)\Delta _{0}$, $\chi =(3g_{s}J/4)\chi _{0}$ and $%
\widetilde{\mu }=\mu +\partial \left\langle H^{\prime }\right\rangle
_{0}/N\partial \delta $. The mean fields $\Delta _{0}$ and $\chi _{0}$ can
be determined by solving these self-consistent equations[\ref%
{hole_concentration}-\ref{dispersion2}]. The SC order parameter is defined as

\begin{equation*}
\Delta _{SC}(\boldsymbol{R}_{ij})=\langle c_{i\uparrow }^{\dag
}c_{j\downarrow }^{\dag }-c_{i\downarrow }^{\dag }c_{j\uparrow }^{\dag
}\rangle
\end{equation*}%
which is related to the variational parameter $\Delta _{0}$ in the
Gutzwiller approximation, \cite{FC Zhang, Vanilla}
\begin{equation}
\Delta _{SC}=g_{t}\Delta _{0}  \label{sop}
\end{equation}

\subsection{Variational Monte Carlo method}

In the VMC calculation, we first rewrite the wavefunction Eq.(\ref%
{wavefunction}) in the Hilbert space with fixed number of $N_{e}$ electrons
doped with even number of $n$ holes,
\begin{equation}
\mid \Psi _{RVB}\rangle =P_{G}(\sum_{\boldsymbol{k}}\frac{v_{\boldsymbol{k}}%
}{u_{\boldsymbol{k}}}c_{\boldsymbol{k}\uparrow }^{\dagger }c_{-\boldsymbol{k}%
\downarrow }^{\dagger })^{(N_{e}-n)/2}\mid 0\rangle  \label{rvb}
\end{equation}%
with%
\begin{eqnarray*}
\frac{v_{\boldsymbol{k}}}{u_{\boldsymbol{k}}} &=&\frac{\Delta _{MC}(%
\boldsymbol{k})}{\epsilon _{MC}(\boldsymbol{k})+\sqrt{\epsilon _{MC}(%
\boldsymbol{k})^{2}+|\Delta _{MC}(\boldsymbol{k})|^{2}}}, \\
\epsilon _{MC}(\boldsymbol{k}) &=&-2t(\cos {k_{x}}+\cos {k_{y}}%
)-4t_{v}^{\prime }\cos {k_{x}}\cos {k_{y}} \\
&&-2t_{v}^{\prime \prime }(\cos {2k_{x}}+\cos {2k_{y}})-\mu _{v}, \\
\Delta _{MC}(\boldsymbol{k}) &=&2\Delta _{v}(cosk_{x}-cosk_{y})
\end{eqnarray*}%
where $\Delta _{v}$ and $\mu _{v}$ are variational parameters, with $\Delta
_{v}$ related to the $d$-wave SC order parameter and $\mu _{v}$ similar to
the chemical potential. Note that we have used subscript $MC$ to distinguish
the parameters here from those adopted in the section B, and that we have
included two additional variational parameters $t_{v}^{\prime }$ and $%
t_{v}^{\prime \prime }$, which are usually not equal to the bare values $%
t^{\prime }$ and $t^{\prime \prime }$ because the constraint strongly
renormalizes the hopping amplitude.\emph{\ }That is to say, the form of $%
\epsilon _{MC}(\boldsymbol{k})$ in the variational wavefunction can be
different from the dispersion function of the non-interacting electrons.
These variational parameters determine the Fermi surface topology. Then, the
quasi-particle excitations are created by adding holes into Eq.(\ref{rvb}):
\begin{equation}
\mid \Psi _{exc}(\boldsymbol{q})\rangle =P_{G}c_{\boldsymbol{q}\uparrow
}^{\dag }(\sum_{\boldsymbol{k}}\frac{v_{\boldsymbol{k}}}{u_{\boldsymbol{k}}}%
c_{\boldsymbol{k}\uparrow }^{\dagger }c_{-\boldsymbol{k}\downarrow
}^{\dagger })^{(N_{e}-n)/2-1}\mid 0\rangle .  \label{rvb1}
\end{equation}%
From Eq.(\ref{rvb1}) we calculate the energy dispersion for a given doping
density by using VMC. The system used in this paper is of $12\times 12$
sites with periodic boundary conditions.\footnotemark[1]  \footnotetext[1]{%
The VMC results reported here are for $H$ in Eq.(\ref{Hamiltonian}) with an
additional term of $(-1/4)n_{i}n_{j}$ in $H_{J}$. This additional term is a
constant at the half-filled.} We then fit the quasi-particle energy with the
formula $a\cdot \sqrt{\varepsilon _{\boldsymbol{k}}^{2}+\Delta _{\boldsymbol{%
k}}^{2}}-b$ to determine the renormalized parameters, with $a$ and $b$ the
fitting parameters. Additionally, in order to eliminate the bias introduced
in the trial wavefunction method, the power-Lanczos method \cite{Y.C. Chen,
C.T. Shih} which is a hybrid of the power and the variational Lanczos method
is used to further improve the trial function. In the power method it can be
easily shown that if a trial wave function $\left\vert \Psi \right\rangle $
is not orthogonal to the ground state, $(\emph{W}-\emph{H})^{\emph{m}%
}\left\vert \Psi \right\rangle $ is proportional to the ground state
wavefunction as the power \emph{m} approaches infinity. \emph{W} is an
appropriately chosen constant to make the ground state energy the largest
eigenvalue of the \emph{W}$-$\emph{H} matrix. In our calculation, the first
order Lanczos method, i.e., $m=1$ is used and the improved trial
wavefunction is $\left\vert PL1\right\rangle =(1+\emph{C}_{1}\emph{H}%
)\left\vert \Psi \right\rangle $. $\emph{C}_{1}$ is a new variational
parameter. The results described below denoted as PL1 are calculated with
the trial wavefunction $(1+\emph{C}_{1}\emph{H})\left\vert \Psi
\right\rangle $.

\section{BASIC PARAMETERS}

\begin{figure}[tbp]
\includegraphics[width=8.0cm,height=8.0cm]{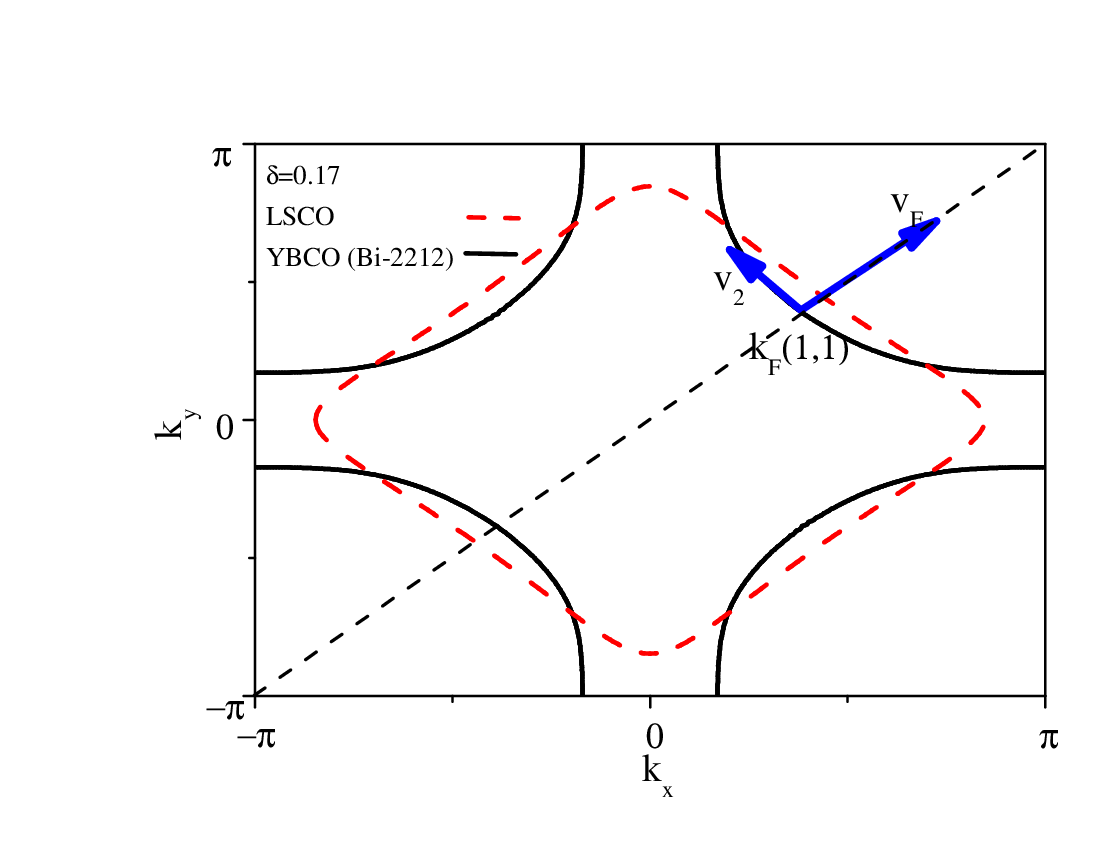}
\caption{(Color online) Illustration of the Fermi surface for LSCO (dashed
square) and for YBCO (solid line), and the location of the gap nodes $%
k_{F}(\pm 1,\pm 1)$. The \textquotedblleft Fermi velocity\textquotedblright\
$v_{F}$ and the \textquotedblleft gap velcoity\textquotedblright\ $v_{2}$
are defined as the slopes of the quasiparticle energy along and
perpendicular to the nodal direction. $v_{F}$ and $v_{2}$ specify the Dirac
cone for the nodal quasiparticle dispersion.}
\label{fs}
\end{figure}

In this section, we discuss the parameters of nodal quasiparticles in the
d-wave SC Cu-oxides and make comparisons between the theory and experiments.
It has been well established in experiments that the cuprate
superconductivity has $d_{x^{2}-y^{2}}$-wave pairing \cite{Tsue}. There are
four nodes in the $\boldsymbol{k}$-space, where the quasiparticle dispersion
$E_{\boldsymbol{k}}=\sqrt{\xi _{\boldsymbol{k}}^{2}+\Delta _{\boldsymbol{k}%
}^{2}}$ approaches zero. The typical Fermi surface (FS) in HTSC
is shown in Fig.\ref{fs}, together with the \textquotedblleft Fermi
velocity\textquotedblright\ $v_{F}$ and the \textquotedblleft gap
velocity\textquotedblright\ $v_{2}$ which are defined as the slopes of $E_{%
\boldsymbol{k}}$ along the direction perpendicular and tangential to the
Fermi surface at the nodes $k_{F}(\pm 1,\pm 1)$. $\sqrt{2}k_{F}$ is the
Fermi wave-vector along the diagonal direction. Different from the
conventional s-wave pairing symmetry, low energy quasiparticles in the
vicinity of these nodes can be easily excited by\ thermal fluctuation,
impurity scattering, or disorder effects. These low energy nodal
quasiparticles predominate physical properties of HTSC at low temperatures.

\subsection{Fermi wave-vector}

\begin{figure}[tbp]
\includegraphics[width=8.5cm,height=6.0cm]{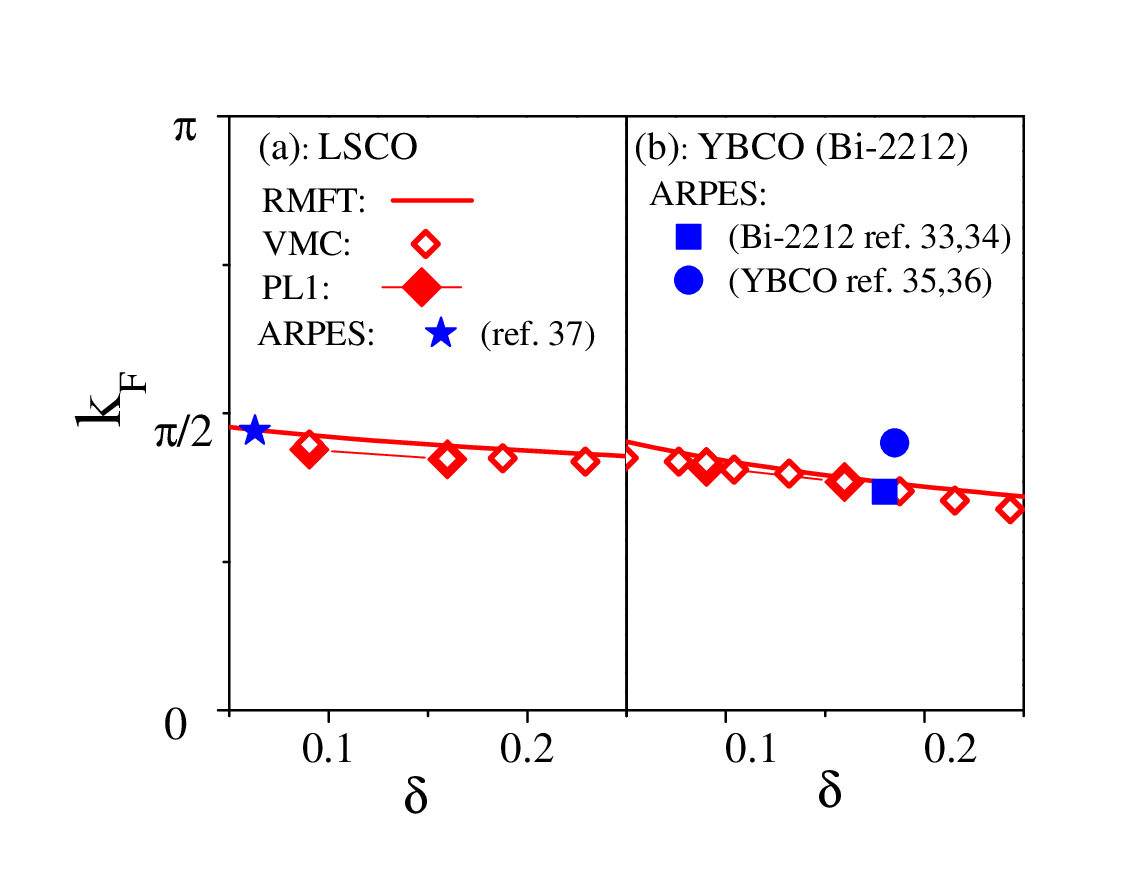}
\caption{(Color online) Comparison of the doping dependence of the Fermi
wave-vector $k_{F}$ obtained from our theoretical calculation with those
obtained by the ARPES\protect\cite{Z.X. Shen6, M. Chiao1, M.C. Schabel, H.
Ding2, J. Mesot} for (a) LSCO and (b) YBCO (Bi-2212). Theoretical results
are obtained using parameters listed in Table I for LSCO and YBCO (Bi-2212)
with the in-plane lattice constant $a=3.8\mathring{A}$. RMFT: renormalized
mean field theory; VMC: variational Monte Carlo; PL1: Power Lanczos to the
first order. Note that, the values from VMC and PL1 may be very close, and
get overlapped with each other in some plots following. }
\label{kf}
\end{figure}

From the ARPES data, it is known that a transition from a hole-like Fermi
surface centered at $(\pi ,\pi )$ to an electron-like Fermi surface centered
at $(0,0)$ occurs slightly above the optimal doping in both LSCO and
Bi-2212, meanwhile the Fermi wave-vector $k_{F}$ shifts just a little bit.
\cite{A. Ino2, D. S. Dessau, T. Yoshida} For Bi-2212, ARPES experiments \cite%
{H. Ding2} suggest that $k_{F}$ is weakly doping-dependent and $\sqrt{2}%
k_{F}\simeq 0.43\mathring{A}^{-1}$ \cite{J. Mesot}. For optimally doped YBCO$%
_{6.95}$, $\sqrt{2}k_{F}\simeq 0.53\mathring{A}^{-1}$ \cite{M. Chiao1, M.C.
Schabel} and for underdoped $La_{2-x}Sr_{x}Cu_{2}O_{4}$ ($x=0.063$), $\sqrt{2%
}k_{F}=0.55\mathring{A}^{-1}$ \cite{Z.X. Shen6} with the lattice constant $%
a=3.8\mathring{A}$. These experimental data are shown in Fig.\ref{kf},
compared with our theoretical calculation where $k_{F}$ is determined by $%
\xi (\boldsymbol{k})=0$ along the diagonal direction. For all the method
considered, we found that $k_{F}$ decreases with increasing doping and $%
k_{F} $ decreases more rapidly in YBCO (Bi-2212) than in LSCO. The values of
$k_{F} $ for the underdoped LSCO and optimally doped Bi-2212 and YBCO agree
qualitatively with the experimental data.

\subsection{Fermi velocity $v_F$}

In the vicinity of the gap nodes, the quasiparticle dispersion can be
expressed as
\begin{equation*}
E(\boldsymbol{k})=\sqrt{v_{F}^{2}\text{ }k_{\parallel
}^{2}+v_{2}^{2}k_{\perp }^{2}}
\end{equation*}%
where $v_{F}$ ($v_{2}$) and $k_{\parallel }$ ($k_{\perp }$) are the
components of the velocity and wave-vector perpendicular (parallel) to the
Fermi surface, respectively.

The Fermi velocity extracted from the slope of the quasiparticle dispersion
obtained by the ARPES is found remarkably universal, independent of the
doping concentration, $v_{F}\approx 250\sim 270km/s$ within an experimental
error of 10-20\%, \cite{J. Mesot, X.J. Zhou}.

\begin{figure}[tbp]
\includegraphics[width=8.5cm,height=6.0cm]{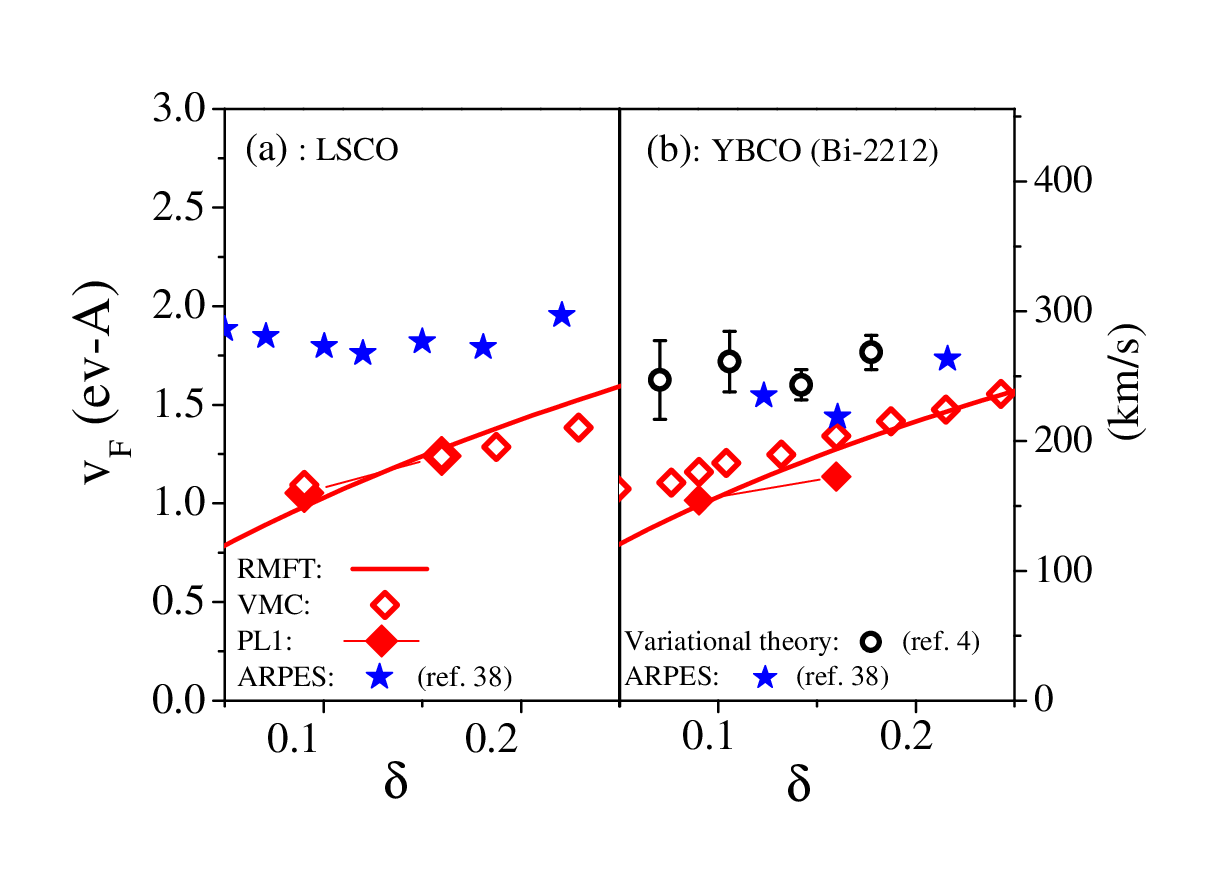}
\caption{(Color online) Fermi velocity $v_{F}$ vs hole concentration $%
\protect\delta $. The results from variational theory in Ref. \protect\cite%
{Randeria} were obtained with the correction of order of $O(J/t)$ included
for the Hubbard model. ($1eV-\mathring{A}\simeq 152km/s$) }
\label{vf}
\end{figure}

In the RMFT, $v_{F}$ and $v_{2}$ are given by the following equations
\begin{eqnarray}
v_{F} &=&\sqrt{2}\left\vert \sin (k_{F})\right\vert   \label{vfvelocity} \\
&&\left\vert 2(g_{t}t+\frac{1}{2}\chi )-4g_{t}|t^{\prime }|\cos
(k_{F})+8g_{t}t^{\prime }\cos (k_{F})\right\vert   \notag \\
v_{2} &=&\left\vert \sqrt{2}\Delta \sin (k_{F})\right\vert
\label{v2velocity}
\end{eqnarray}%
Shown in Fig.\ref{vf} is the value of $v_{F}$ obtained from Eq.[\ref%
{vfvelocity}]. $v_{F}$ increases with doping. The VMC gives essentially the
same result. In the optimally-doped and overdoped regimes, this trend does
not deviate greatly from the experimentally observed universality. However,
the value of $v_{F}$ appears underestimated in the RMFT for the extended t-J
model compared with the experimental data and with that obtained for the
Hubbard model by including the correction of order of $O(J/t)$ reported
previously by Paramekanti et al. \cite{Vanilla, Randeria}.

\subsection{\textquotedblleft Gap velocity\textquotedblright\ $v_{2}$}

The \textquotedblleft gap velocity\textquotedblright\ $v_{2}$ is the slope
of the SC energy gap along the Fermi surface at the gap node. Together with $%
k_{F}$ and $v_{F}$, $v_{2}$ specifies the Dirac cone for the nodal
quasiparticle dispersion. Among them $v_{2}$ plays a crucial role in
determining the nodal physics of HTSC. This is because $k_{F}$ and $v_{F}$
are rather universal, depending weakly on the doping concentration.
Furthermore, $v_{2}$ is much smaller than $v_{F}$, and many physical
properties are related to $v_{2}$ in the form of the ratio $v_{F}/v_{2}$,
therefore a small variation in $v_{2}$ may lead to a drastic change of $%
v_{F}/v_{2}$, hence of some physical quantities.

Experimentally, $v_{2}$ is difficult to be determined accurately. It depends
strongly on the doping concentration and other material properties. A number
of experiments may be used to extract $v_{2}$. These experiments include
ARPES \cite{J. Mesot, A. Ino2}, the temperature dependence of in-plane
magnetic penetration depth $\lambda (T)$ \cite{C. Panagopoulos1}, the
electronic specific heat $C_{el}$ \cite{D.A. Wright, K.A. Moler, A. Junod, H
H Wen}, and the linear residual thermal conductivity $\kappa _{0}/T|_{T=0}$
\cite{M. Chiao1, M. Chiao2, M. Sutherland, J. Takeya}. The linear residual
thermal conductivity is robust against renormalization due to quasiparticle
interactions and vertex corrections. In the SC state, $\kappa
_{0}/T|_{T=0}\varpropto v_{F}/v_{2}+v_{2}/v_{F}$ is universal \cite{P.A.
Lee2} and it does not depend on the impurity scattering rate.

\begin{figure}[tbp]
\includegraphics[width=8.5cm,height=10.0cm]{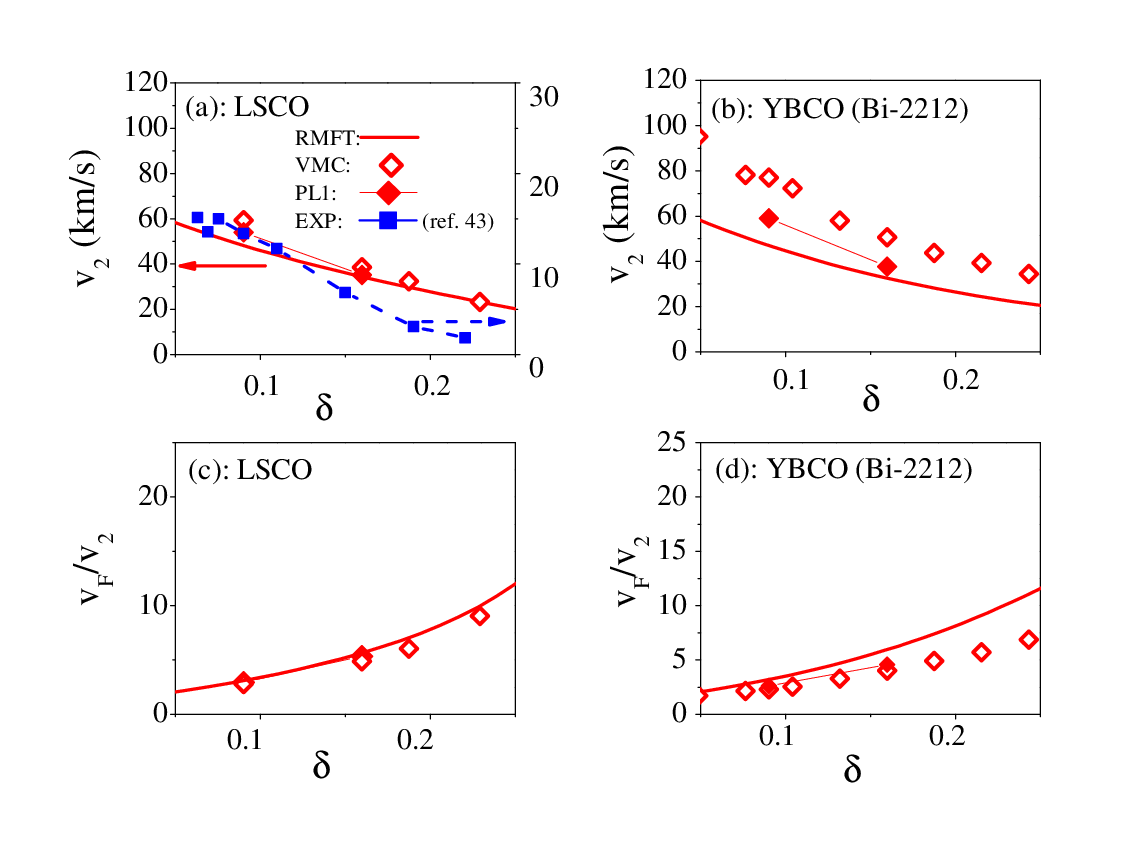}
\caption{(Color online) \textit{Panels (a,b) }: \textquotedblleft Gap
velocity\textquotedblright\ $v_{2}$ vs hole concentration $\protect\delta $.
For optimal doped YBCO (Bi-2212), $v_{2}=10\sim 20km/s$ was reported by
various kinds of experiments \protect\cite{M. Chiao1, M. Sutherland, J.
Mesot}. The experimental data indicated in panel (a) is achieved by
measuring the magnetic field dependence of the specific heat on LSCO at the
zero temperature limit \protect\cite{H H Wen}. \textit{Panels (c,d):} the
ratio $v_{F}/v_{2}$ vs hole concentration $\protect\delta $. The doping
dependence of the ratio $v_{F}/v_{2}$ is similar to that observed in the
thermal conductivity experiments (see Fig.\protect\ref{heat}). }
\label{v2}
\end{figure}

Fig.\ref{v2} shows the doping dependence of $v_{2}$ and the ratio $%
v_{F}/v_{2}$. The calculation shows that $v_{2}$ drops quasi-linearly with
increasing doping. This behavior is consistent with the recent experiment of
the magnetic field dependence of the specific heat on LSCO \cite{H H Wen}.
Our calculation gives $v_{2}=20\sim 30km/s$ at an optimal doping, which is
larger than the experimentally reported value of $10\sim 20km/s$ for YBCO
(Bi-2212) obtained from the thermal conductivity measurement \cite{M.
Sutherland, J. Takeya} and ARPES \cite{J. Mesot, M. Chiao1}, and of $7km/s$
for LSCO obtained from the measurements of magnetic field dependence of the
specific heat at the zero temperature limit \cite{H H Wen}. In the RMFT, $%
v_{F}/v_{2}=1$ at zero doping. As doping increases, $v_{F}/v_{2}$ increases
rapidly. The doping dependence is qualitatively consistent with the data
reported in the thermal conductivity measurements. \cite{M. Sutherland, J.
Takeya}

In our theoretical calculation, despite the great difference between the
hopping integrals $t^{\prime }$ and $t^{\prime \prime }$ for YBCO (bi-2212)
and LSCO, the values of $v_{2}$ are only slightly different. This result is
also in qualitative agreement with the thermal conductivity measurements. In
the next section we will use our theoretical result of $v_{2}$ and $v_{F}$
to extract some physical observables and compare with experimental results.

\section{NODAL PHYSICS}

In HTSC, $d$-wave pairing symmetry leads to a dome-like quasiparticle
dispersion around the gap nodes. In the SC state, the gapless quasiparticle
excitations in the vicinity of nodes dominate low temperature physical
properties. It is of fundamental importance to explore physical properties
of these quasiparticle excitations.

In the clean limit, the density of states (DOS), $\rho (\omega )$, of low
lying quasiparticles near the nodes is linear,
\begin{equation}
\rho (\omega )=\frac{2}{\pi }\frac{1}{v_{F}v_{2}\text{ }}\omega  \label{dos}
\end{equation}%
The linear coefficient of $\rho (\omega )$ is inversely proportional to the
nodal velocities $v_{F}$ and $v_{2}$. This linearity in energy of $\rho
(\omega )$ leads to many unconventional physical behaviors such as the
quadratic electronic specific heat \cite{M. Chiao1, T. Xiang, D.A. Wright,
J. Mesot}, the linear residual thermal conductivity \cite{M. Chiao1, M.
Chiao2, M. Sutherland, M. Sutherland-addition} and the linear decreasing of
superfluid density \cite{P.A. Lee, W.N. Hardy}. Experimental observations of
these behaviors have provided some of the early evidences for unconventional
$d_{x^{2}-y^{2}}$ pairing symmetry in HTSC. The nature of the interactions
of nodal quasiparticles is not so clear in HSTC. \cite{T. Xiang, J. Mesot}
Some have used a renormalization factor to describe the effect of
quasiparticle interactions on the electronic specific heat and on the
in-plane magnetic penetration depth. \cite{T. Xiang, J. Mesot} In this
paper, we shall neglect quasiparticle interactions and set the
renormalization factor to unity.

\subsection{Electronic specific heat}

\begin{figure}[tbp]
\includegraphics[width=8.5cm,height=6.0cm]{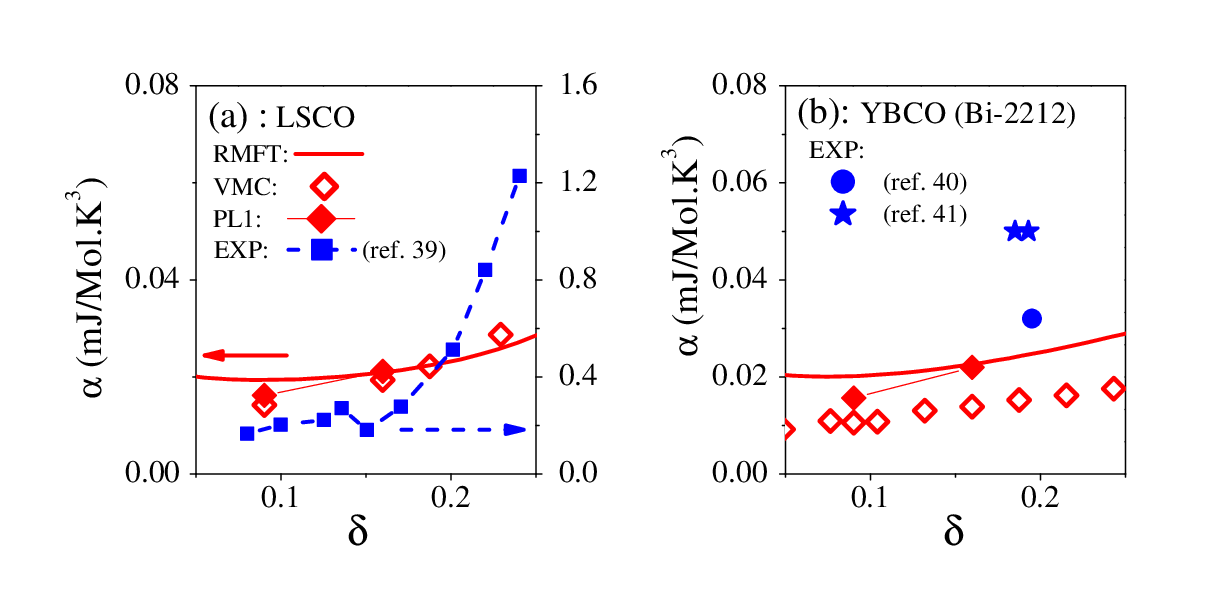}
\caption{(Color online) The quadratic coefficient of the electronic specific
heat: $\protect\alpha =C_{el}/T^{2}$ vs hole concentration $\protect\delta $%
. }
\label{specificheat1}
\end{figure}

The linear low energy DOS $\rho (\omega )$ leads to a quadratic temperature
dependence of the low temperature electronic specific heat in the HTSC,
given by%
\begin{equation}
C_{el}=\gamma T=\alpha T^{2},\alpha =\frac{21.6}{\pi }\frac{k_{B}^{3}}{\hbar
^{2}}\frac{1}{v_{F}v_{2}\text{ }}  \label{cel}
\end{equation}%
Fig.\ref{specificheat1} compares our theoretical results with the
experiments for LSCO and YBCO. \cite{D.A. Wright, K.A. Moler, C.
Panagopoulos1} The experimental result for LSCO shows a general tendency to
increase as doping increases, and the rapid increase of $\alpha $ in the
overdoped region might be due to the Fermi level crossing of the flat band
at $(\pi ,0)$, which yields an additional channel to thermally excited
quasiparticles. In our theoretical results, the doping dependence of $\alpha
$ is similar for LSCO and YBCO. The values of $\alpha $ are about $0.01\sim
0.03mJ/Mol\cdot K^{3}$ which are comparable to the experimental value of
YBCO, \cite{D.A. Wright, K.A. Moler} but much smaller than the value of LSCO
\cite{C. Panagopoulos1}.

\subsection{Thermal conductivity}

In the presence of a small amount of disorder or impurities, the nodal
quasiparticles are delocalized and can carry both heat and charge. For
dilute non-magnetic impurities, there will be a residual normal fluid due to
these delocalized and conductive quasiparticles. The most striking property
of this conduction mechanism is the universal limit, i.e. the quasiparticle
transport is independent of the scattering rate as $T\rightarrow 0$. With
increasing the impurity concentration, the mean free path is reduced, but
the normal fluid density increases. \cite{P.A. Lee1, P.A. Lee2} In the SC
state with a random distribution of impurities of an energy scale $%
E_{im}<k_{B}T_{c}$, the low temperature thermal conductivity is linear, \cite%
{P.A. Lee1, P.A. Lee2} and is given by
\begin{equation}
\frac{\kappa _{0}}{T}|_{T=0}=\frac{k_{B}^{2}}{3\hbar }\frac{n}{d}(\frac{v_{F}%
}{v_{2}}+\frac{v_{2}}{v_{F}})  \label{heatconduction}
\end{equation}%
where $d/n$, the stacking distance between two nearest neighboring $CuO_{2}$
planes, has the values of $6.6\mathring{A}$, $5.8\mathring{A}$, $7.72%
\mathring{A}$ for LSCO, YBCO, Bi-2212, respectively. This formula is
obtained within the self-consistent $T$-matrix approximation, and it may
break down if the impurity scattering gets too strong. \cite{P.A. Lee3} This
universal behavior of the thermal conductivity provides a robust and direct
measurement of $v_{F}/v_{2}$ in the SC state.

\begin{figure}[tbp]
\includegraphics[width=8.5cm,height=8.5cm]{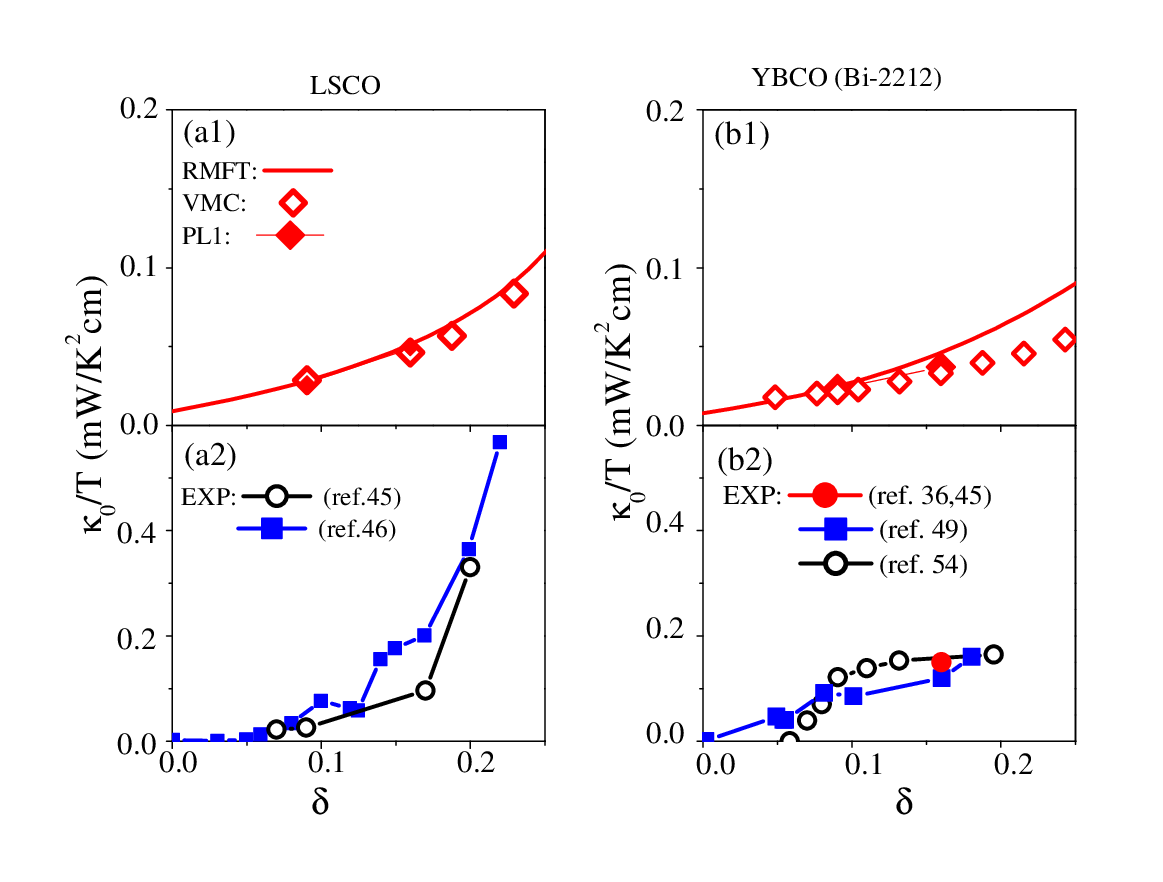}
\caption{(Color online) Linear residual thermal conductivity $\protect\kappa %
_{0}/T$ vs hole concentration $\protect\delta $. }
\label{heat}
\end{figure}

Fig.\ref{heat} shows our theoretical results of $\kappa _{0}/T|_{T=0}$
compared with the experimental results for LSCO and YBCO (Bi-2212). \cite{M.
Chiao1, M. Sutherland, J. Takeya, X. F. Sun-addition, M. Sutherland-addition}
Experimentally, above a critical doping $\delta _{pc}$, both LSCO and YBCO
(Bi-2212)\ are thermal metals and $\kappa _{0}/T|_{T=0}$ increases steadily
as $\delta $ in the underdoped regime and very rapidly in the overdoped
regime. Such observation strongly supports the notion that there are well
defined nodal quasiparticles in the clean limit. The difference of the
residual thermal conductivity between LSCO and YBCO (Bi-2212) is much
smaller compared with the electronic specific heat shown in Fig.\ref%
{specificheat1}. In the lightly underdoped regime $\delta <\delta _{pc}$,
the low temperature behavior of $\kappa _{0}/T$ remains unclear \cite{X. F.
Sun-addition, M. Sutherland-addition}. However, it is clear that $\kappa
_{0}/T\rightarrow 0$ as $T\rightarrow 0$ in LSCO\cite{N.E. Hussey, D. G.
Hawthorn}. The thermal insulating behavior in LSCO is probably due to the
localization of quasiparticles due to disorder effects.

In our calculation, the SC state and the delocalized quasiparticles are
assumed to prevail in the heavily underdoped region. The theoretical results
deviate from the experimental ones by a factor of $2\sim 4$. We attribute
this discrepancy to the overestimated gap velocity\ $v_{2}$ in the theory.

\subsection{In-plane magnetic penetration depth}

The magnetic penetration depth $\lambda (T)$ is related to the superfluid
density $\rho _{s}$ by
\begin{equation*}
\frac{\rho _{s}(T)}{m^{\ast }}=\frac{\rho _{s}(0)}{m^{\ast }}-\frac{\rho
_{n}(T)}{m^{\ast }}=\frac{c^{2}}{4\pi e^{2}}\frac{1}{\lambda ^{2}}
\end{equation*}%
where $m^{\ast }$\ is the effective mass of the charge carriers, and will be
assumed to be doping independent, and $\rho _{n}$ is the normal fluid
density. \cite{P.A. Lee} At low temperatures, $\rho _{n}$ is contributed
from the thermally excited quasiparticles near nodes, and can be given by
\begin{equation*}
\frac{\rho _{n}(T)}{m^{\ast }}=\left( \frac{2\ln 2}{\pi }\right) \frac{%
n\,v_{F}}{d\,v_{2}}\frac{k_{B}T}{\hbar ^{2}}.
\end{equation*}%
The linear temperature coefficient of $\rho _{s}(T)/m^{\ast }$ is
proportional to $v_{F}/v_{2}$.

At low temperatures, the temperature dependence of $\lambda (T)$ is very
weak, and $\lambda (0)$ is about several thousands angstroms. \cite{C.
Panagopoulos, C. Panagopoulos1, W.N. Hardy, T. Xiang3, M.R. Trunin, G.
Lamura} The first and second derivative of the penetration depth with
respect to temperature can be approximately expressed as%
\begin{eqnarray}
\frac{d\lambda (T)}{dT}|_{T\rightarrow 0} &=&\lambda ^{3}(0)4\ln 2\frac{e^{2}%
}{c^{2}}\frac{k_{B}}{\hbar ^{2}}\frac{n}{d}\frac{v_{F}}{v_{2}},
\label{penetrationdepth1} \\
\frac{d}{dT}\lambda ^{-2}|_{T\rightarrow 0} &=&-8\ln 2\frac{e^{2}}{c^{2}}%
\frac{k_{B}}{\hbar ^{2}}\frac{n}{d}\frac{v_{F}}{v_{2}}.
\label{penetrationdepth2}
\end{eqnarray}

\begin{figure}[tbp]
\includegraphics[width=8.5cm,height=9.0cm]{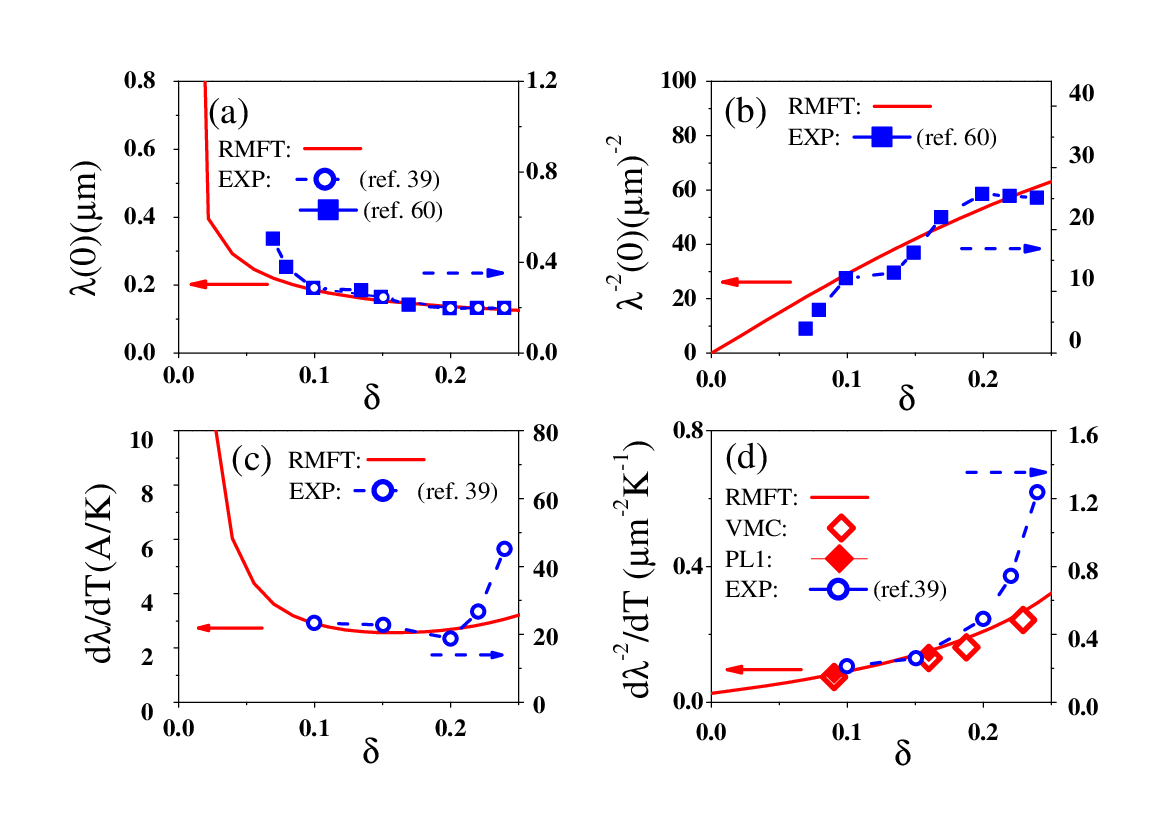}
\caption{(Color online) In-plane magnetic penetration depth in LSCO vs hole
concentration $\protect\delta $. }
\label{lscopenetration}
\end{figure}

\begin{figure}[tbp]
\includegraphics[width=8.5cm,height=9.0cm]{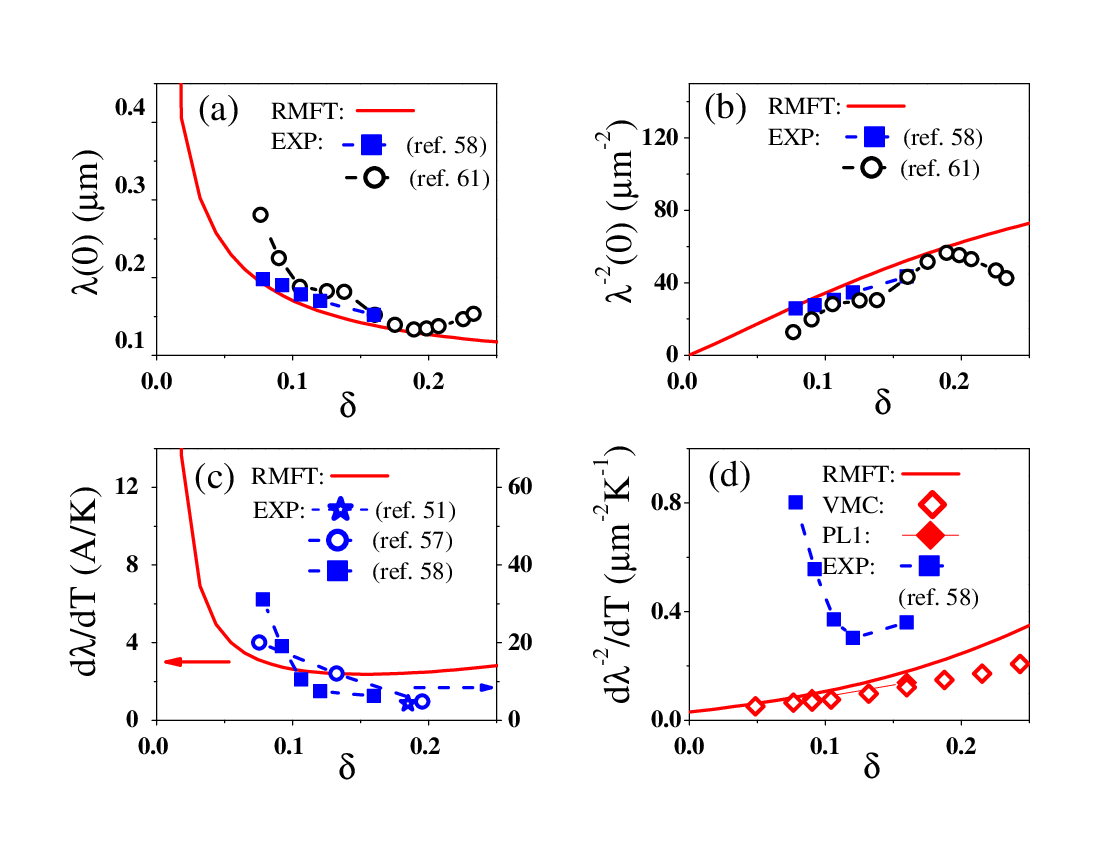}
\caption{(Color online) In-plane magnetic penetration depth in YBCO vs hole
concentration $\protect\delta $. }
\label{ybcopenetration}
\end{figure}

\begin{figure}[tbp]
\includegraphics[width=8.5cm,height=9.0cm]{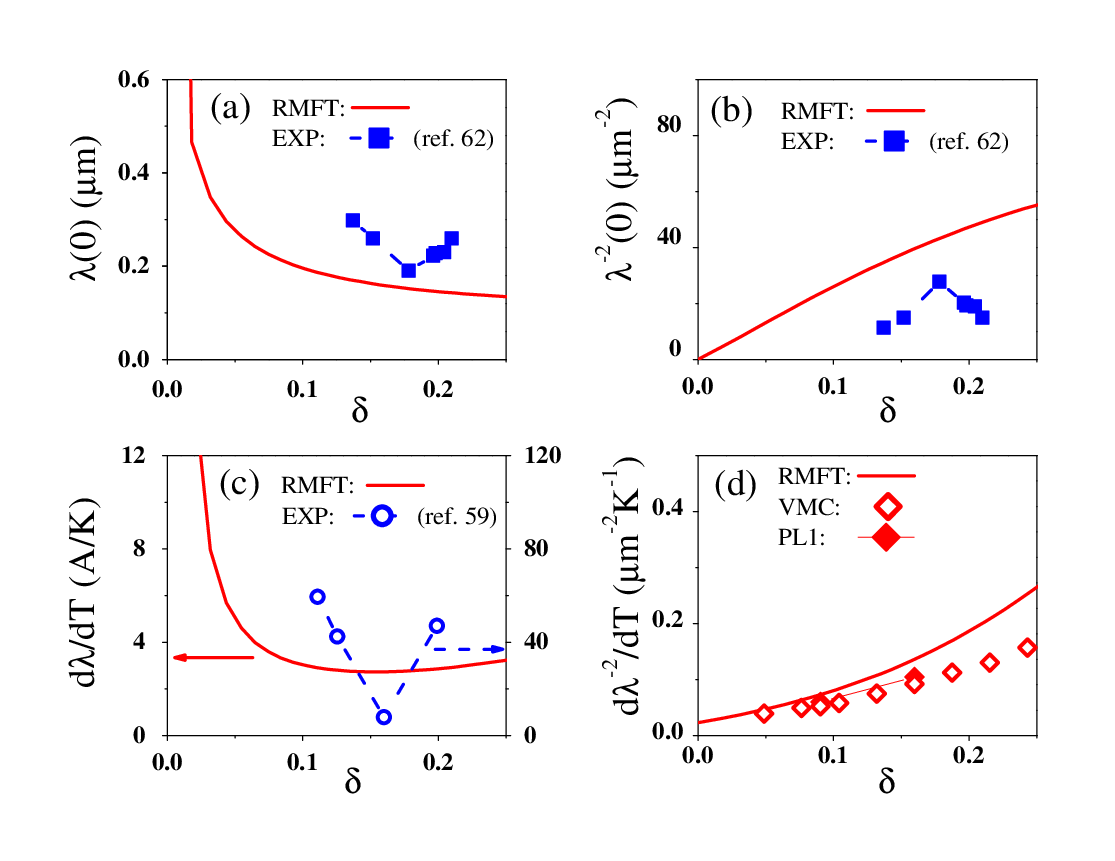}
\caption{(Color online) In-plane magnetic penetration depth in $%
Bi_{2}Sr_{2}Ca_{1-x}Y_{x}Cu_{2}O_{8+\protect\delta }$ vs hole concentration $%
\protect\delta $. }
\label{bsccopenetration}
\end{figure}

Panels (a) and (b) of Fig.\ref{lscopenetration}-\ref{bsccopenetration} show
the zero temperature in-plane magnetic penetration depth $\lambda (0)$ and $%
\lambda ^{-2}(0)$. Experimentally, as $\delta $ increases, $\lambda (0)$ in
LSCO monotonically decreases \cite{C. Panagopoulos, C. Panagopoulos1}, while
$\lambda (0)$ in YBCO and in $Bi_{2}Sr_{2}Ca_{1-x}Y_{x}Cu_{2}O_{8+\delta }$
increases with doping in the overdoped region \cite{C. Bernhard, G. Villard}%
. The experimental results of $\lambda ^{-2}(0)$ in LSCO and underdoped YBCO
(Bi-2212) show a linear doping dependence, supporting the idea that the zero
temperature superfluid density $\rho _{s}(0)$ is proportional to the doping
concentration in the underdoped region. In our RMFT, in the SC phase, $%
\lambda ^{-2}(0)$ is nearly linear with the hole doping and $\lambda (0)$
diverges at zero doping within the approximation that all optical spectral
weights are condensed to the zero energy in the t-J model. Our theoretical
results of $\lambda (0)(\lambda ^{-2}(0))$ agree with the experimental data
for LSCO and YBCO in the underdoped region. In $%
Bi_{2}Sr_{2}Ca_{1-x}Y_{x}Cu_{2}O_{8+\delta }$, our theoretical results show
a discrepancy with the experiments.

Panels (c,d) of Fig.\ref{lscopenetration}-\ref{bsccopenetration} show the
derivatives of the penetration depth with respect to temperature, $d\lambda
(T)/dT$ and $d\lambda ^{-2}/dT$. In the underdoped or slightly overdoped
region, $d\lambda (T)/dT$\ decreases with increasing doping in all three
compounds. \cite{C. Panagopoulos, C. Panagopoulos1, W.N. Hardy, T. Xiang3,
M.R. Trunin, G. Lamura} In the heavily overdoped region $d\lambda (T)/dT$
increases with doping in both LSCO and $Bi_{2}Sr_{2}Ca_{1-x}Y_{x}Cu_{2}O_{8+%
\delta }$. In LSCO, $d\lambda ^{-2}/dT$ increases steadily with doping.\cite%
{C. Panagopoulos, C. Panagopoulos1} In YBCO, an opposite tendency was
observed \cite{M.R. Trunin} in the underdoped region $\delta <0.10$. The
anomalous increase in the underdoped region was previously shown to
qualitatively agree with the behavior resulted from the d-density wave
state. \cite{M.R. Trunin, Q.H. Wang, S. Tewari} In our RMFT, similar doping
dependence of $d\lambda (T)/dT$ is obtained. However, there is a great
discrepancy on the absolute values between the experiments and our
theoretical results. We argue that some other mechanisms may be responsible
for the large value of $d\lambda (T)/dT$ observed in experiments. \cite{G.
Blumberg1}

\section{OTHER PHYSICAL PROPERTIES}

\subsection{Drude Weight}

The Drude weight or the superfluid stiffness $D$ is a measurement of
superfluid condensation. In the linear-response theory, within the
approximation that in the t-J model all optical spectral weights are
condensed to zero energy, the Drude weight $D$ can be given by\cite{E.
Dagotto}
\begin{equation}
D=\left( \frac{e^{2}}{4\pi \varepsilon _{0}\hbar ^{2}}\right) ^{-1}\left(
\frac{n}{d}\right) ^{-1}\frac{2}{\pi }\int_{0}^{\infty }d\omega \func{Re}%
\sigma (\omega )=-\frac{\left\langle E_{kin}\right\rangle }{2}
\label{drude1}
\end{equation}%
$D$ is related to the plasma frequency $\omega _{p}^{\ast }$ by $\omega
_{p}^{2}/8=\int_{0}^{\infty }d\omega \func{Re}\sigma (\omega )$. In optical
reflectivity measurements, the frequency-dependent conductivities can be
derived from the reflectivity spectra. By integrating the spectral weight
below $1.25eV$, $(\omega _{p}^{\ast })^{2}$ was found to vanish linearly
with the decrease of doping concentration in the low doping regime, and for
optimally doped YBCO $(\hbar \omega _{p}^{\ast })^{2}\simeq 4.5eV$ along the
a-axis (without the contribution from the $CuO$ chain) i.e. $D\simeq 145meV$%
. \cite{S. L. Cooper}

\begin{figure}[tbp]
\includegraphics[width=8.5cm,height=5.0cm]{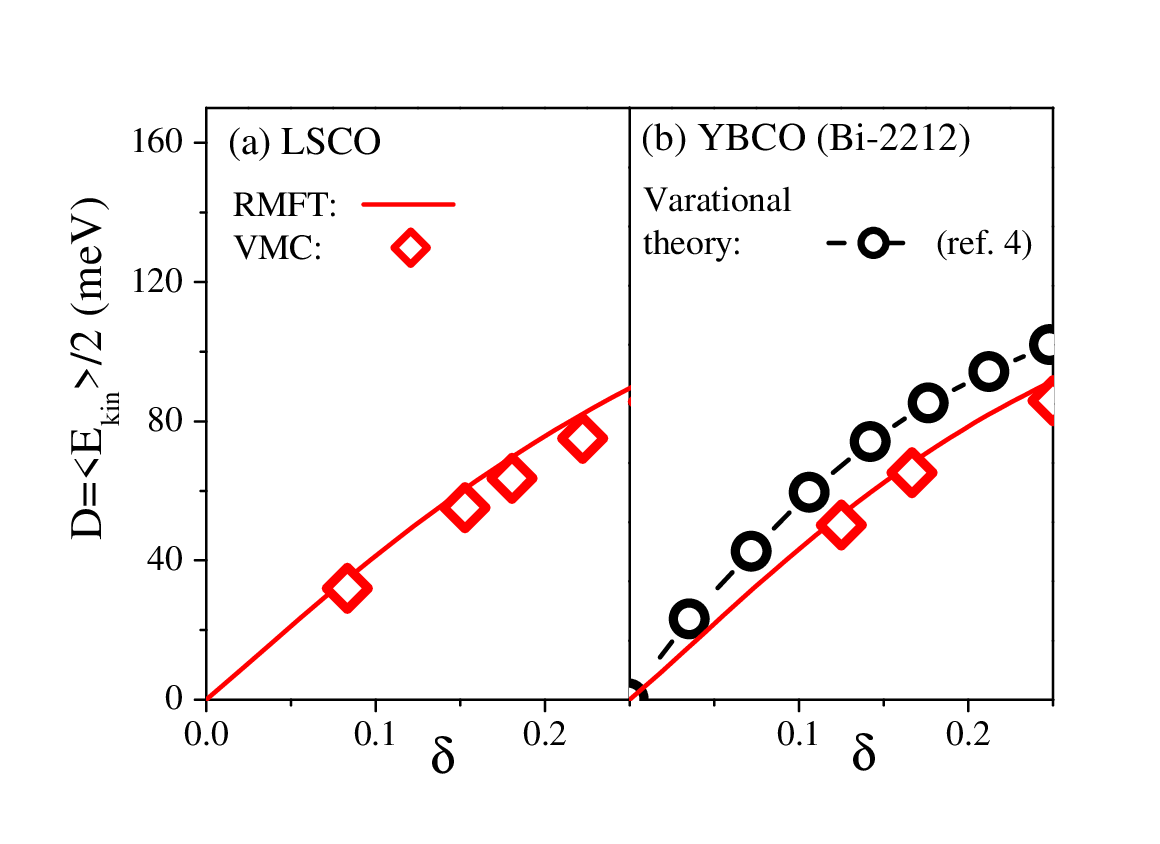}
\caption{(Color online) Drude weight $D$ vs hole concentration $\protect%
\delta $. The results from the variational theory given in Ref. \protect\cite%
{Randeria} are for the Hubbard model with a finite cut-off to get rid of the
contribution due to transitions from the ground state to the
\textquotedblleft upper Hubbard band\textquotedblright . In optical
reflectivity measurements, the Drude weight is proportional to $(\protect%
\omega _{p}^{\ast })^{2}$, for optimally doped YBCO $(\hbar \protect\omega %
_{p}^{\ast })^{2}\simeq 4.5eV$ along the a-axis i.e. $D\simeq 145meV$.
\protect\cite{S. L. Cooper}}
\label{Drude}
\end{figure}

Fig.\ref{Drude} shows the RMFT and VMC results for Drude weight. Our results
agree with those obtained with a finite cut-off of the integration in Eq.\ref%
{drude1} to get rid of the contributions due to transitions from the ground
state to the \textquotedblleft upper Hubbard band\textquotedblright ,\cite%
{Randeria} those results include the correction of order of $O(J/t)$. The
Drude weight increases almost linearly in the underdoped regime. Around the
optimal doping, our results of the Drude weight is about $60meV$, in
agreement with the optical reflectivity experimental data given in Ref.\cite%
{S. L. Cooper}.

\subsection{The antinodal quasiparticle energy $E(\protect\pi ,0)$}

Around the antinodal point $(\pi ,0)$, the quasiparticle dispersion becomes
flat. This flat band has been studied intensively by experiments. \cite{A.
Fujimori, A. Ino2, Z.X. Shen4, K. Gofron, D. S. Dessau} In the RMFT study,
the energy of quasiparticle excitations at $(\pi ,0)$ is given by:
\begin{equation}
E(\pi ,0)=\sqrt{(-4g_{t}t^{\prime }-4g_{t}t^{\prime \prime }-\widetilde{\mu }%
)^{2}+4\Delta ^{2}}.  \label{renormlaizedepi0}
\end{equation}

\begin{figure}[tbp]
\includegraphics[width=8.5cm,height=6.0cm]{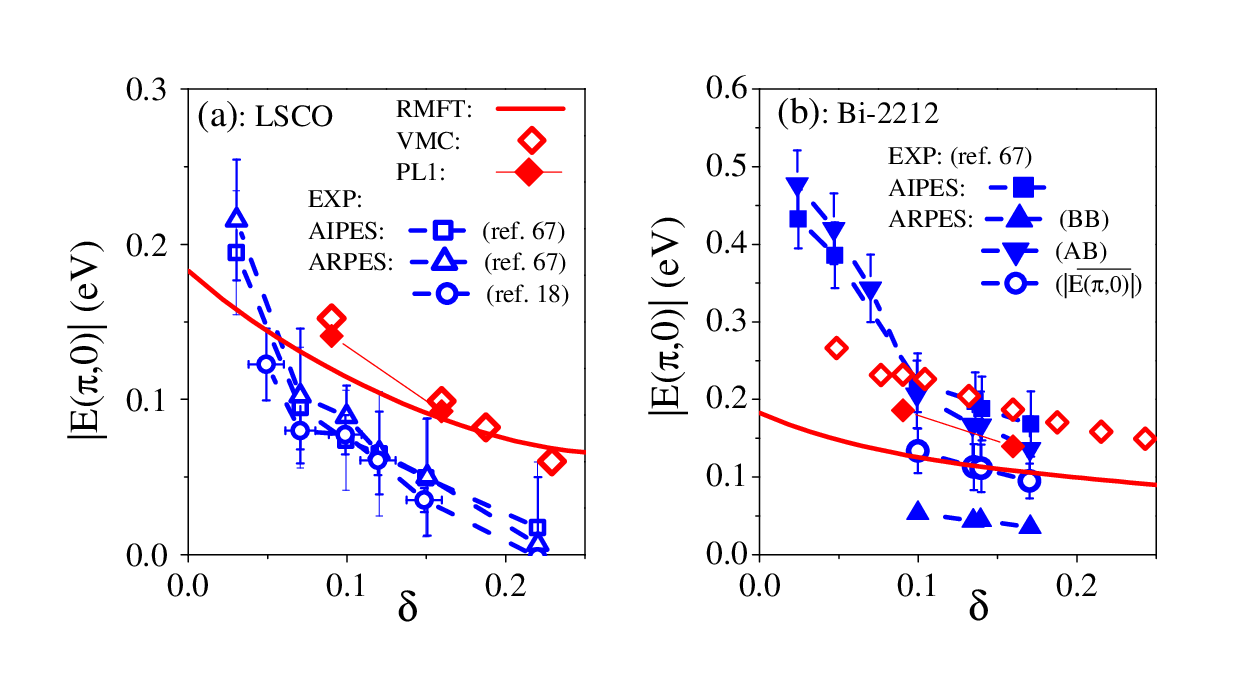}
\caption{(Color online) The quasiparticle energy $E(\protect\pi ,0)$ vs hole
concentration $\protect\delta $. }
\label{epi0}
\end{figure}

Fig.\ref{epi0} shows the doping dependence of $E(\pi ,0)$ obtained in our
calculation compared with the experimental results\cite{A. Fujimori, A. Ino2}%
. The experimental results obtained by AIPES (angle-integrated photoemission
spectroscopy) and ARPES agree well with each other. In LSCO, the energy
position of the flat band lies about $200\sim 300meV$ below the Fermi energy
for lightly underdoped state, and is shifted up to the Fermi level quickly
with increasing doping, finally crosses the Fermi level at optimal doping or
slightly overdoping. In Bi-2212, two branches of flat bands (bonding band
and anti-bonding band) were observed due to the bilayer splitting. They are
determined by the low and high binding energies of the peak-dip-hump
character. The average $\overline{E(\pi ,0)}$ of the two bands\cite{A.
Fujimori} is shown in Fig.\ref{epi0}(b). The bonding band has the same
doping dependence as that in LSCO. The anti-bonding band lies much higher
than the bonding band.

In our calculation, without taking the bilayer coupling into consideration,
our theoretical calculation of $E(\pi ,0)$ in Bi-2212\ should correspond to
the average $\overline{E(\pi ,0)}$. Similar doping dependence is observed
experimentally. $E(\pi ,0)$ approaches to the Fermi level with increasing
doping, but does not get very close to the Fermi level even in the overdoped
regime in contrary to LSCO. In VMC and PL1 simulation, the value of $E(\pi
,0)$ is much more closer to $\overline{E(\pi ,0)}$ of Bi-2212 but is much
larger than in LSCO.

\subsection{Chemical potential shift}

\begin{figure}[tbp]
\includegraphics[width=8.5cm,height=6.0cm]{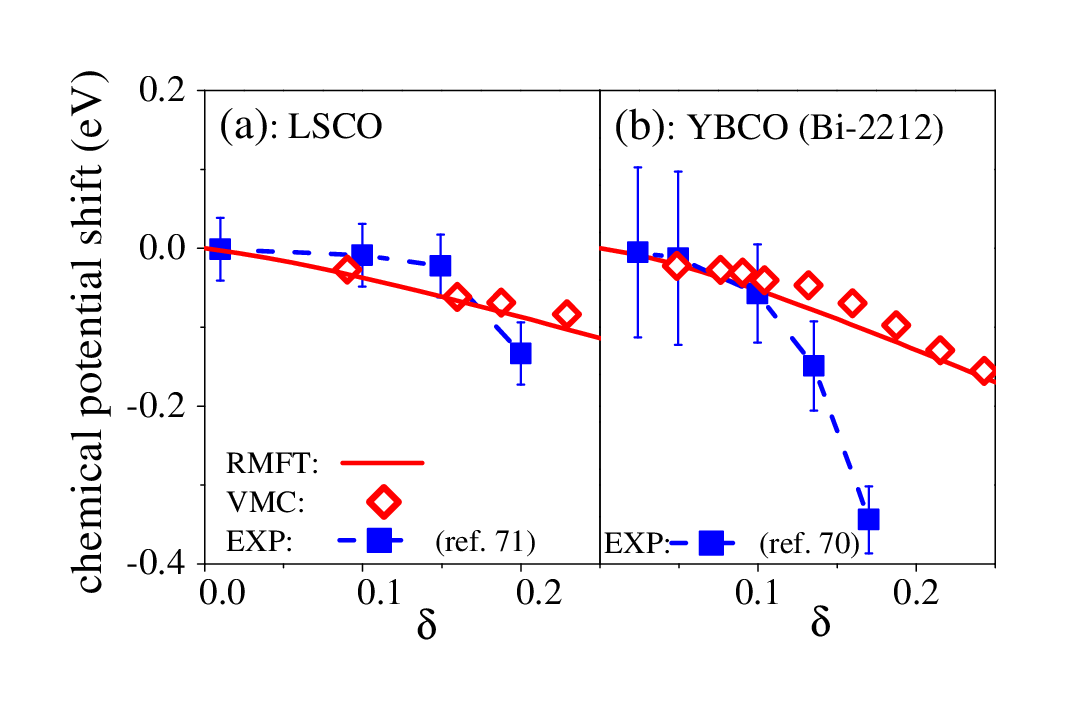}
\caption{(Color online) Chemical potential shift $\widetilde{\protect\mu }$
vs hole concentration $\protect\delta $. The experimental data were deduced
from the shifts of photoemission and inverse-photoemission spectra of the
core states of LSCO \protect\cite{A. Ino} and Bi-2212 \protect\cite{A.
Fujimori2}. }
\label{cpshift}
\end{figure}

Fig.\ref{cpshift} shows the electron chemical potential shift $\widetilde{%
\mu }$, compared with experimental data for LSCO \cite{A. Ino} and Bi-2212
\cite{A. Fujimori2}. In the RMFT, $\widetilde{\mu }$ is given by
\begin{equation}
\widetilde{\mu }=\mu +\frac{1}{N}\frac{\partial \left\langle H^{\prime
}\right\rangle _{0}}{\partial \delta }  \label{chemical potential}
\end{equation}%
The experimental data were deduced from the shifts of photoemission and
inverse-photoemission spectra of the core states of LSCO and Bi-2212. In
LSCO the chemical potential shift $\widetilde{\mu }$ was found to be pinned
close to zero energy in underdoped regime. \cite{A. Fujimori2, A. Ino, K.
Yamada-addition} In Bi-2212, the chemical potential shift is not pinned at
zero energy and shows a more rigid-band-like behavior.

In our calculations, the chemical potential shift agrees qualitatively with
the experimental data. It is also consistent with the result obtained by the
exact diagonalization of the t-t'-J model \cite{S. Maekawa}. Furthermore,
the shift is found to be larger in Bi-2212 than in LSCO in the entire hole
doping range, in agreement with the experiments.

\subsection{Quasiparticle spectral weight}

\begin{figure}[tbp]
\includegraphics[width=8.5cm,height=6.0cm]{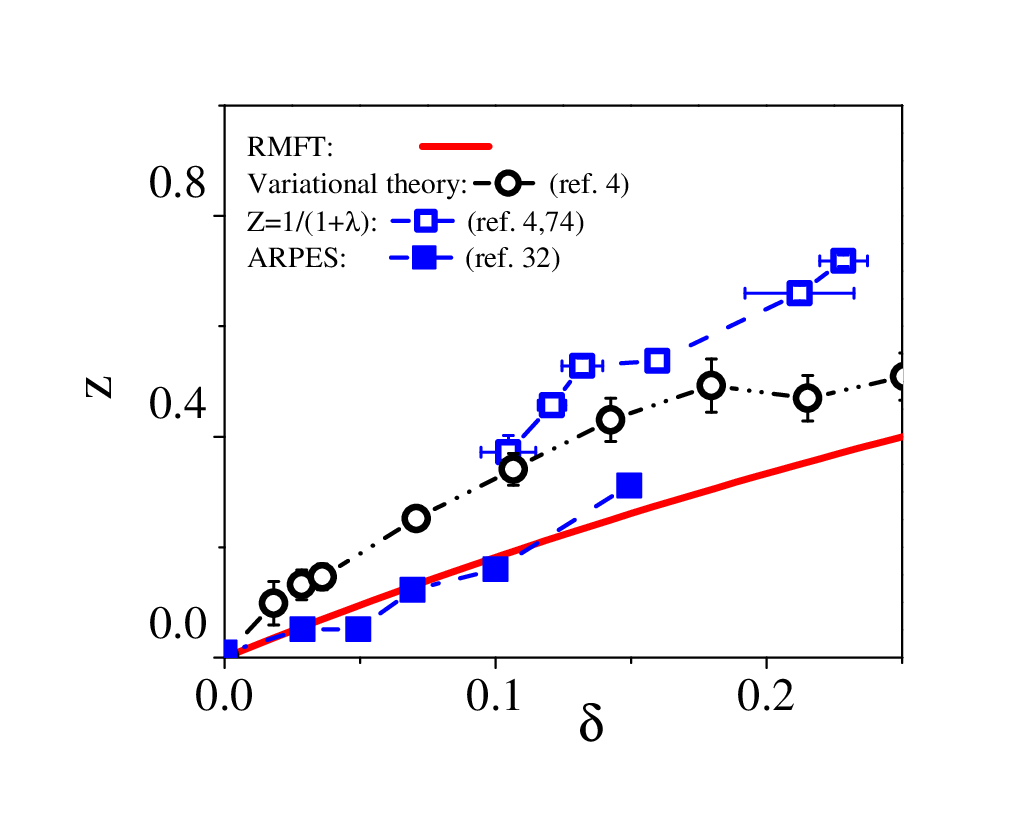}
\caption{(Color online) Nodal quasiparticle weight $Z$ vs hole concentration
$\protect\delta $. $Z=1/(1+\protect\lambda )$ was estimated in ref.%
\protect\cite{Randeria}, where $\protect\lambda $ is the coupling constant
estimated from the spectra function in ARPES \protect\cite{P.D. Johnson}.
The results from VMC simulation \protect\cite{Randeria} is presented for a
complementary comparison. }
\label{za}
\end{figure}

Fig.\ref{za} shows the nodal quasiparticle spectral weight $Z$. In ARPES
experiments the quasiparticle spectral weight $Z$ can be deduced from the
spectral weight of the quasiparticle coherent peak at the gap-nodes \cite{T.
Yoshida} or from the formula $Z=1/(1+\lambda )$, \cite{Randeria} where the
coupling constant $\lambda $ can be extracted from the real part of
self-energy $\func{Re}\Sigma (\boldsymbol{k},\omega )$ of the spectral
function \cite{P.D. Johnson}. In the RMFT analysis, the nodal quasiparticle
spectral weight is equal to the renormalized factor of the hopping term $%
g_{t}$. Our theoretical results for the doping dependence of $Z$ agree well
with the experimental results. The nodal quasiparticle spectral weight grows
almost linearly in the whole doping region as shown. The results from VMC
\cite{Randeria} are presented for a complementary comparison.

\subsection{Superconducting gap}

Experimentally the maximal superconducting gap $\Delta _{m}$ can be measured
by the thermal conductivity, ARPES, or other techniques. For example, from
the thermal conductivity \cite{M. Sutherland}, $\Delta _{m}$ can be
determined by assuming $\Delta _{m}=\hbar k_{F}v_{2}/2$ with
\textquotedblleft universal\textquotedblright\ Fermi velocity $v_{F}$ \cite%
{X.J. Zhou} and \textquotedblleft weakly\textquotedblright\ doping dependent
$k_{F}$ \cite{H. Ding2}. In ARPES, the midpoint shift of the leading edge of
the quasiparticle spectral at $(\pi ,0)$ is approximately equal to $\Delta
_{m}$. One can also determine $\Delta _{m}$ by fitting the gap dispersion on
the Fermi surface with the formula $\Delta (\phi )=\Delta _{m}\cos 2\phi $,
where $\phi $ is the Fermi surface angle. \cite{M. R. Norman, P. J. White,
J. C. Campuzano, H. Ding5, A. Ino}

\begin{figure}[tbp]
\includegraphics[width=8.5cm,height=9.0cm]{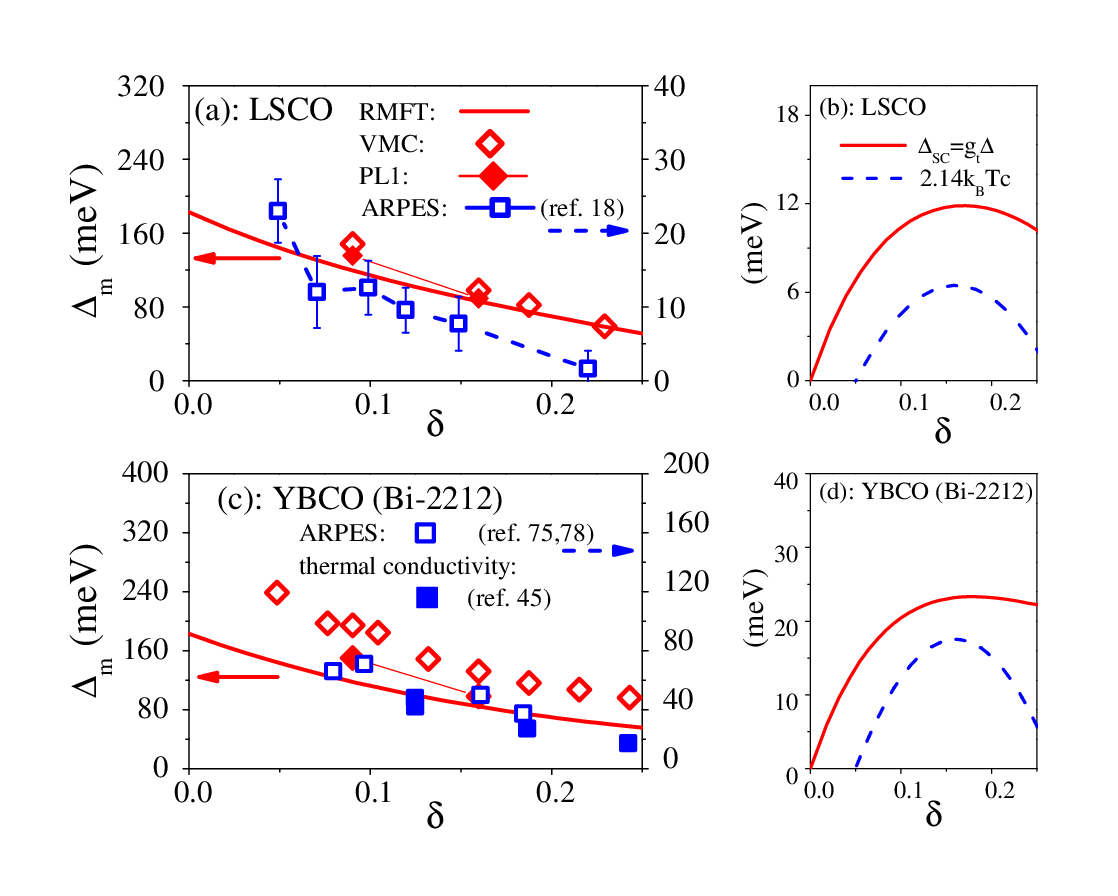}
\caption{(Color online) \textit{Panels (a,c): }The maximal superconducting
gap $\Delta _{m}$ vs hole concentration $\protect\delta $. \textit{Panels
(b,d)}: comparison of the superconducting order parameter $\Delta
_{SC}=g_{t}\Delta $ in our RMFT calculation with the gap $\Delta
_{BCS}=2.14k_{B}T_{c}$, where $T_{c}$ is estimated from $T_{c}=T_{c}^{\max
}(1-82.6(\protect\delta -0.16)^{2})$ \protect\cite{M.R. Presland} ($%
T_{c}^{\max }=35K$ for LSCO and $T_{c}^{\max }=95K$ for YBCO (Bi-2212)). }
\label{delta}
\end{figure}

In Fig.\ref{delta} (a,c), our theoretical results of $\Delta _{m}=\Delta
(\cos k_{x}-\cos k_{y})|_{(\pi ,0)}$ are shown and compared with the
experimental data \cite{A. Ino, M. Sutherland, M. R. Norman, P. J. White}.
The doping dependence of $\Delta _{m}$ agrees with the experiments, but the
absolute values are about twice larger than the experimental ones in YBCO
(Bi-2212).

Fig.\ref{delta} (b,d) compare the value of $\Delta _{SC}=g_{t}\Delta $ with
the BCS gap $\Delta _{BCS}\simeq 2.14k_{B}T_{c}$ obtained by assuming $%
T_{c}=T_{c}^{\max }(1-82.6(\delta -0.16)^{2})$, ($T_{c}^{\max }=95K$ for
Bi-2212, $35K$ for LSCO)\cite{M.R. Presland}. $\Delta _{SC}$ and $\Delta
_{BCS}$ are roughly proportional to each other.

\section{Summary}

In this paper we have made a systematic comparison between the plain vanilla
RVB theory and a broad spectrum of experimental data of low energy physical
properties in cuprates. In our theoretical calculations with both RMFT and
VMC, the only parameters are the spin coupling and the hopping integrals of
electrons on the $CuO_{2}$ plane, which are known quite accurately. We have
found a qualitatively good agreement between the theory and the experiments
on almost all the quantities we have studied, including the specific heat,
the thermal conductivity, the in-plane magnetic penetration depth, and the
antinodal quasiparticle energy, the Drude weight, and the superconducting
gap. The agreement on the doping dependences of these properties is
remarkable except in the heavy overdoped regime. The major discrepancy is on
the absolute values of some quantities, which may be attributed to the large
value of $v_{2}$ estimated in the theory. The comparison would be quite
satisfactory quantitatively if one had used a theoretical value of $v_{2}$
by $2\sim 4$ times smaller, which indicates a possibility of overestimate of
the gap. (It is known that the gap estimated in the VMC calculation is
overestimated by a factor of 2 or more). This discrepancy could also be due
to the simplification of the model Hamiltonian or the approximate
wavefunction. More investigation will be needed to address these issues.

\section{ACKNOWLEDGEMENT}

This work was partially supported by the RGC grant in Hong Kong, and NSC
94-2112-M-001-003 grant in Taiwan.


\begin{thebibliography}{99}
\bibitem{Vanilla} P. W. Anderson, P. A. Lee, M. Randeria, T. M. Rice, N.
Triveni, and F. C. Zhang, J. Phys.: Condens. Matter 16, R755 (2004).

\bibitem{anderson87} P. W. Anderson, Science \textbf{235}, 1196 (1987).

\bibitem{FC Zhang} F C Zhang, C Gros, T M Rice and H Shiba, Supercond. Sci.
Technol., 1, 36-46 (1988), and cond-mat/0311604.

\bibitem{Randeria} A. Paramekanti, M. Randeria, and N. Trivedi, Phys. Rev.
Lett. 87, 217002 (2001); M. Randeria, A. Paramekanti, and N. Trivedi, Phys.
Rev. B69, 144509 (2004); A. Paramekanti, M. Randeria, and Nandini Trivedi,
Phys. Rev. B70, 054504 (2004).

\bibitem{Zhang-Rice} F. C. Zhang and T. M. Rice, Phys. Rev. B37, R3759
(1988).

\bibitem{Gros89} C. Gros, Phys. Rev. B38, R931 (1988); C. Gros, Ann. Phys.
189, 53 (1989); H. Yokoyama and H. Shiba, J. Phys. Soc. Japan 57, 2482
(1988).

\bibitem{TKLee} T. K. Lee, C. T. Shih, Y. C. Chen, and H. Q. Lin, Phys. Rev.
Lett. 89, 279702 (2002); M. Ogata and A. Himeda, J. Phys. Soc. Japan 72, 374
(2003).

\bibitem{CTShih} C. T. Shih, T. K. Lee, R. Eder, C.-Y. Mou, and Y. C. Chen,
Phys. Rev. Lett. 92, 227002 (2004); C. T. Shih, Y. C. Chen, C. P. Chou, and
T. K. Lee, Phys. Rev. B70, 220502(R) (2004).

\bibitem{Kotliar} G. Kotliar and J. Liu, Phys. Rev. B38, R5142 (1988); Y.
Suzumura, Y. Hasegawa, and H. Fukuyama, J. Phys. Soc. Japan 57, 2768 (1988).
P. A. Lee, Phys. Rev. Lett. 63, 680 (1989).

\bibitem{anderson-ong} P. W. Anderson and N. P. Ong, cond-mat/0405518; M.
Randeria, R. Sensarma, N. Trivedi, and F. C. Zhang, Phys. Rev. Lett. 95,
137001 (2005).

\bibitem{Zhang03} F. C. Zhang, Phys. Rev. Lett. 90, 207002 (2003); G.
Baskaran, Phys. Rev. Lett. 90, 197007 (2003); J. Y. Gan, Y. Chen, Z. B. Su,
and F. C. Zhang, Phys. Rev. Lett. 94, 067005 (2005); J. Liu, J. Schmalian,
N. Trivedi, Phys. Rev. Lett. 94, 127003 (2005); see also R. Laughlin,
cond-mat/0209269.

\bibitem{BEdegger} B. Edegger, V. N. Muthukumar, C. Gros, and P. W.
Anderson, cond-mat/0512646.

\bibitem{Rice06} K.-Y. Yang, T. M. Rice, and F.-C Zhang, cond-mat/0602164.

\bibitem{TYoshida} T. Yoshida, X. J. Zhou, K. Tanaka, W. L. Yang, Z.
Hussain, Z.-X. Shen, A. Fujimori, S. Komiya, Y. Ando, H. Eisaki, T.
Kakeshita, and S. Uchida, cond-mat/0510608.

\bibitem{Y.C. Chen} Y.C. Chen and T.K. Lee, Phys. Rev. B 51, 6723 (1995).

\bibitem{C.T. Shih} C.T. Shih, Y.C. Chen, and T.K. Lee, Phys. Rev. B 57, 627
(1998).

\bibitem{E. Pavarini} E. Pavarini, I. Dasgupta, T. Saha-Dasgupta, O. Jepsen,
O. K. Anderson, Phys. Rev. Lett. 87, 047003 (2001).

\bibitem{A. Ino2} A. Ino, C. Kim, M. Nakamura, T. Yoshida, T. Mizokawa, A.
Fujimori, Z.-X. Shen, T. Kakeshita, H. Eisaki, and S. Uchida, Phys. Rev. B
65, 094504 (2002).

\bibitem{Z.X. Shen} A. Damascelli, Zhi-Xun Shen, Z. Hussain, Rev. Mod. Phys.
75, 473, (2003).

\bibitem{K. B. Lyons} K. B. Lyons, P. A. Fleury, L. F. Schneemeyer, and J.
V. Waszczak, Phys. Rev. Lett. 60, 732 (1988).

\bibitem{J. M. Tranquada} J. M. Tranquada, G. Shirane, B. Keimer, S.
Shamoto, and M. Sato, Phys. Rev. B 40, 4503 (1989).

\bibitem{R. Coldea} R. Coldea, S. M. Hayden, G. Aeppli, T. G. Perring, C. D.
Frost, T. E. Mason, S.-W. Cheong, and Z. Fisk, Phys. Rev. Lett. 86, 5377
(2001).

\bibitem{S. M. Hayden} S. M. Hayden, G. Aeppli, P. Dai, H. A. Mook, T. G.
Perring, S.-W. Cheong, Z. Fisk, F. Dogan, and T. E. Mason, Physica B
241-243, 765 (1998).

\bibitem{G. Blumberg} G. Blumberg, M. Kang, M. V. Klein, K. Kadowaki, and C.
Kendziora, Science 278, 1427 (1997).

\bibitem{S. Sugai} Shunji Sugai, Shin--ichi Shamoto and Masatoshi Sato,
Phys. Rev. B 38, 6436 (1988).

\bibitem{G. Blumberg1} G. Blumberg, P. Abbamonte, M. V. Klein, W. C. Lee,
and D. M. Ginsberg, L. L. Miller, A. Zibold, Phys. Rev. B 53, R11930 (1996).

\bibitem{Gutzwiller} M. C. Gutzwiller, Phys. Rev. Lett. 10, 159 (1963).

\bibitem{Brinkman-rice} W. Brinkman and T. M. Rice, Phys. Rev. B2 (1970).

\bibitem{Vollhardt} D. Vollhardt, Rev. Mod Phys. 56, 99 (1984).

\bibitem{Tsue} C. C. Tsuei and J. R. Kirtley, Rev. Mod. Phys, 72, 969 (2000).

\bibitem{D. S. Dessau} D. S. Dessau, Z.-X. Shen, D. M. King, D. S. Marshall,
L. W. Lombardo, P. H. Dickinson, A. G. Loeser, J. DiCarlo, C.-H Park, A.
Kapitulnik, and W. E. Spicer, Phys. Rev. Lett. 71, 2781 (1993).

\bibitem{T. Yoshida} T. Yoshida, X. J. Zhou, T. Sasagawa, W. L. Yang, P. V.
Bogdanov, A. Lanzara, Z. Hussain, T. Mizokawa, A. Fujimori, H. Eisaki, Z.-X.
Shen, T. Kakeshita, S. Uchida, Phys. Rev. Lett. 91, 027001 (2003).

\bibitem{H. Ding2} H. Ding, M. R. Norman, T. Yokoya, T. Takeuchi, M.
Randeria, J. C. Campuzano, T. Takahashi, T. Mochiku, and K. Kadowaki, Phys.
Rev. Lett., 78, 2628 (1997).

\bibitem{J. Mesot} J. Mesot, M. R. Norman, H. Ding, M. Randeria, J. C.
Campuzano, A. Paramekanti, H. M. Fretwell, A. Kaminski, T. Takeuchi, T.
Yokoya, T. Sato, T. Takahashi, T. Mochiku, and K. Kadowaki, Phys. Rev. Lett.
83, 840, (1999).

\bibitem{M.C. Schabel} M. C. Schabel, C.-H. Park, A. Matsuura, Z.-X. Shen,
D. A. Bonn, Ruixing Liang, and W. N. Hardy, Phys. Reb. B 57, 6090 (1998).

\bibitem{M. Chiao1} May Chiao, R. W. Hill, Christian Lupien, Louis
Taillefer, P. Lambert, R. Gagnon, and P. Fournier , Phys. Rev. B, 62, 3554
(2000).

\bibitem{Z.X. Shen6} X. J. Zhou, Junren Shi, T. Yoshida, T. Cuk, W. L. Yang,
V. Brouet, J. Nakamura, N. Mannella, Seiki Komiya, Yoichi Ando, F. Zhou, W.
X. Ti, J. W. Xiong, Z. X. Zhao, T. Sasagawa, T. Kakeshita, H. Eisaki, S.
Uchida, A. Fujimori, Zhenyu Zhang, E. W. Plummer, R. B. Laughlin, Z.
Hussain, Z.-X. Shen, cond-mat/0405130.

\bibitem{X.J. Zhou} X. J. Zhou, T. Yoshida, A. Lanzara, P. V. Bogdanov, S.
A. Kellar, K. M. Shen, W. L. Yang, F. Ronning, T. Sasagawa, T. Kakeshita, T.
Noda, H. Eisaki, S. Uchida, C. T. Lin, F. Zhou, J. W. Xiong, W. X. Di, Z. X.
Zhao, A. Fujimori, Z. Hussain, Z. X. Shen, Nature 423, 398 (2003).

\bibitem{C. Panagopoulos1} C. Panagopoulos, B. D. Rainford, J. R. Cooper, W.
Lo, J. L. Tallon, J. W. Loram, J. Betouras, Y. S. Wang, and C. W. Chu, Phys.
Rev. B 60, 14617 (1999).

\bibitem{D.A. Wright} D. A. Wright, J. P. Emerson, B. F. Woodfield, J. E.
Gordon, R. A. Fisher, and N. E. Phillips, Phys. Rev. Lett. 82, 1550 (1999).

\bibitem{K.A. Moler} Kathryn A. Moler, David L. Sisson, Jeffrey S. Urbach,
Malcolm R. Beasley, Aharon Kapitulnik, D. J. Baar, R. Liang, and W. N.
Hardy, Phys. Rev. B, 55, 3954 (1997).

\bibitem{A. Junod} A. Junod, B. Revaz, Y. Wang, and A. Erb., Physica B
284-288, 1043 (2000).

\bibitem{H H Wen} Hai-Hu Wen, Lei Shan, Xiao-Gang Wen, Yue Wang, Hong Gao,
Zhi-Yong Liu, Fang Zhou, Jiwu Xiong, and Wenxin Ti, Phys. Rev. B 72, 134507
(2005).

\bibitem{M. Chiao2} May Chiao, R. W. Hill, Christian Lupien, Bojana Popic,
Robert Gagnon, and Louis Taillefer, Phys. Rev. Lett., 82, 2943 (1999).

\bibitem{M. Sutherland} Mike Sutherland, D. G. Hawthorn, R. W. Hill, F.
Ronning, S. Wakimoto, H. Zhang, C. Proust, Etienne Boaknin, C. Lupien, Louis
Taillefer, Ruixing Liang, D. A. Bonn, W. N. Hardy, Robert Gagnon, N. E.
Hussey, T. Kimura, M. Nohara, and H. Takagi, Phys. Rev. B 67, 174520 (2003).

\bibitem{J. Takeya} J. Takeya, Yoichi Ando, Seiki Komiya, and X. F. Sun,
Phys. Rev. Lett. 88, 077001 (2002).

\bibitem{P.A. Lee2} P. A. Lee, Phys. Rev. Lett. 71, 1887 (1993).

\bibitem{T. Xiang} T. Xiang, C. Panagopoulos, Phys. Rev. B, 61, 6343 (2000).

\bibitem{M. Sutherland-addition} M. Sutherland, S. Y. Li, D. G. Hawthorn, R.
W. Hill, F. Ronning, M. A. Tanatar, J. Paglione, H. Zhang, Louis Taillefer,
J. DeBenedictis, Ruixing Liang, D. A. Bonn, and W. N. Hardy, Phys. Rev.
Lett. 94, 147004 (2005).

\bibitem{P.A. Lee} P. A. Lee and Xiao-Gang Wen, Phys. Rev. Lett. 78, 4111
(1997).

\bibitem{W.N. Hardy} W. N. Hardy, D. A. Bonn, D. C. Morgan, Ruixing Liang,
and Kuan Zhang, Phys. Rev. Lett. 70, 3999 (1993).

\bibitem{P.A. Lee1} A. C. Durst, and P. A. Lee, Phys. Rev. B 62, 1270 (2000).

\bibitem{P.A. Lee3} P. A. Lee, N. Ngaosa, and Xiao-Gang Wen, Rev. Mod. Phys.
78, 17 (2006).

\bibitem{X. F. Sun-addition} X. F. Sun, Kouji Segawa, and Yoichi Ando, Phys.
Rev. B 72, 100502(R) (2005).

\bibitem{N.E. Hussey} N. E. Hussey, S. Nakamae, K. Behnia, H. Takagi, C.
Urano, S. Adachi, S. Tajima, Phys. Rev. Lett. 85, 4140 (2000).

\bibitem{D. G. Hawthorn} D. G. Hawthorn, R. W. Hill, C. Proust, F. Ronning,
Mike Sutherland, Etienne Boaknin, C. Lupien, M. A. Tanatar, Johnpierre
Paglione, S. Wakimoto, H. Zhang, Louis Taillefer, T. Kimura, M. Nohara, H.
Takagi, and N. E. Hussey, Phys. Rev. Lett. 90, 197004 (2003).

\bibitem{T. Xiang3} C. Panagopoulos, J. R. Cooper, and T. Xiang, Phys. Rev.
B, 57, 13422 (1998).

\bibitem{M.R. Trunin} M. R. Trunin, Yu. A. Nefyodov, and A. F. Shevchun,
Phys. Rev. Lett. 92, 067006 (2004).

\bibitem{G. Lamura} G. Lamura, J. Le Cochec, A. Gauzzi, F. Licci, D. Di
Castro, A. Bianconi, and J. Bok, Phys. Rev. B 67, 144518 (2003).

\bibitem{C. Panagopoulos} C. Panagopoulos, T. Xiang, W. Anukool, J. R.
Cooper, Y. S. Wang, and C. W. Chu, Phys. Rev. B 67, 220502(R) (2003).

\bibitem{C. Bernhard} C. Bernhard, J. L. Tallon, Th. Blasius, A. Golnik, and
Ch. Niedermayer , Phys. Rev. Lett. 86, 1614 (2001).

\bibitem{G. Villard} G. Villard, D. Pelloquin, and A. Maignan, Phys. Rev. B
58, 15231 (1998).

\bibitem{Q.H. Wang} Qiang-Hua Wang, Jung Hoon Han, and Dung-Hai Lee, Phys.
Rev. Lett. 87, 077004 (2001).

\bibitem{S. Tewari} Sumanta Tewari, Hae-Yong Kee, Chetan Nayak, and Sudip
Chakravarty, Phys. Rev. B 64, 224516 (2000).

\bibitem{E. Dagotto} E. Dagotto, Rev. Mod. Phys. 66, 763 (1994).

\bibitem{S. L. Cooper} S. L. Cooper, D. Reznik, A. Kotz, M. A. Karlow, R.
Liu, M. V. Klein, W. C. Lee, J. Giapintzakis, and D. M. Ginsberg, B. W.
Veal, and A. P. Paulikas, Phys. Rev. B 47, 8233 (1993).

\bibitem{A. Fujimori} K. Tanaka, T. Yoshida, A. Fujimori, D.H. Lu, Z.-X.
Shen, X.-J. Zhou, H. Eisaki, Z. Hussain, S. Uchida, Y. Aiura, K. Ono, T.
Sugaya, T. Mizuno, and I. Terasaki, Phys. Rev. B 70, 092503 (2004).

\bibitem{Z.X. Shen4} D. S. Marshall, D. S. Dessau, A. G. Loeser, C-H. Park,
A. Y. Matsuura, J. N. Eckstein, I. Bozovic, P. Fournier, A. Kapitulnik, W.
E. Spicer, and Z.-X. Shen, Phys. Rev. Lett., 76, 4841 (1996).

\bibitem{K. Gofron} K. Gofron, J. C. Campuzano, A. A. Abrikosov, M.
Lindroos, A. Bansil, H. Ding, D. Koelling, and B. Dabrowski, Phys. Rev.
Lett. 73, 3302 (1994).

\bibitem{A. Fujimori2} N. Harima, A. Fujimori, T. Sugaya, and I. Terasaki,
Phys. Rev. B 67, 172501 (2003).

\bibitem{A. Ino} A. Ino, T. Mizokawa, and A. Fujimori, K. Tamasaku, H.
Eisaki, S. Uchida, T. Kimura, T. Sasagawa, and K. Kishio, Phys. Rev. Lett.
79, 2101 (1997).

\bibitem{K. Yamada-addition} K. Yamada, C. H. Lee, K. Kurahashi, J. Wada, S.
Wakimoto, S. Ueki, H. Kimura, Y. Endoh, S. Hosoya, G. Shirane, R. J.
Birgeneau, M. Greven, M.A. Kastner, and Y. J. Kim, Phys. Rev. B 57, 6165
(1998).

\bibitem{S. Maekawa} T. Tohyama and S. Maekawa, Phys. Rev. B 67, 092509
(2003).

\bibitem{P.D. Johnson} P. D. Johnson, T. Valla, A. V. Fedorov, Z. Yusof, B.
O. Wells, Q. Li, A. R. Moodenbaugh, G. D. Gu, N. Koshizuka, C. Kendziora,
Sha Jian, and D. G. Hinks, Phys. Rev. lett. 87, 177007 (2001).


\bibitem{M. R. Norman} M. R. Norman, H. Ding, M. Randeria, J. C. Campuzano,
T. Yokoya, T. Takeuchi, T. Takahashi, T. Mochiku, K. Kadowaki, P.
Guptasarma, and D. G. Hinks, Nature 392, 157 (1998).

\bibitem{H. Ding5} H. Ding, M. R. Norman, J. C. Campuzano, M. Randeria, A.
F. Bellman, T. Yokoya and T. Takahashi, T. Mochiku, and K. Kadowaki, Phys.
Rev. B 54, R9678 (1996).

\bibitem{J. C. Campuzano} J. C. Campuzano, H. Ding, M. R. Norman, H. M.
Fretwell, M. Randeria, A. Kaminski, J. Mesot, T. Takeuchi, T. Sato, T.
Yokoya, T. Takahashi, T. Mochiku, K. Kadowaki, P. Guptasarma, D. G. Hinks,
Z. Konstantinovic, Z. Z. Li, and H. Raffy, Phys. Rev. Lett. 83, 3709 (1999).

\bibitem{P. J. White} P. J. White, Z.-X. Shen, C. Kim, J. M. Harris, A. G.
Loeser, P. Fournier, and A. Kapitulnik, Phys. Rev. B 54, R15669 (1996).

\bibitem{M.R. Presland} M. R. Presland et al., Physica C 176, 95 (1991)
\end{thebibliography}
\end{document}